\documentclass[a4paper,fleqn,usenatbib]{mnras}

\usepackage{newtxtext,newtxmath}

\usepackage[T1]{fontenc}
\usepackage{ae,aecompl}

\usepackage{graphicx}	
\usepackage{amsmath}	
\usepackage{amssymb}	
\usepackage[usenames,dvipsnames]{xcolor}

\usepackage{gensymb}    
\usepackage{latexsym}
\usepackage{amsfonts}
\usepackage{epstopdf}
\usepackage{rotating}
\usepackage{pdflscape}
\usepackage{changepage}
\usepackage{url}
\usepackage{color}
\usepackage{footnote}
\usepackage{cleveref}
\crefname{section}{§}{§§}
\Crefname{section}{§}{§§}
\usepackage{float}
\usepackage{subfig}
\usepackage{caption}
\usepackage{booktabs}
\usepackage{fixltx2e}
\usepackage{threeparttable}
\usepackage{longtable}
\usepackage{supertabular,booktabs}
\usepackage{floatpag}
\makesavenoteenv{itemize}
\usepackage{booktabs}
\usepackage{xspace}
\usepackage{color, colortbl}
\usepackage{listings}
\usepackage{enumerate}
\usepackage{multirow}
\usepackage{fixltx2e}
\usepackage{tikz}
\usepackage{diagbox}
\usepackage{multirow}
\usepackage{cellspace}
\usepackage{makecell}
\newcolumntype{x}[1]{>{\centering\arraybackslash}p{#1}}
\usepackage{tikz}

\newcommand{\gaia}{\textit{Gaia}\xspace}

\newcommand{\porb}{\mbox{$P_\mathrm{orb}$}\xspace}
\newcommand{\pc}{\mbox{$\mathrm{pc}^{-3}$}\xspace}

\usepackage{soul}

\title[The 150 pc CV sample from \textit{Gaia} DR2]{
A Volume Limited Sample of Cataclysmic Variables from \textit{Gaia} DR2: Space Density and Population Properties}

\author[A. F. Pala et al.]{A.~F.~Pala,$^{1,2}$\thanks{E-mail: apala@eso.org}, 
B.~T.~G\"ansicke,$^{2}$ 
E.~Breedt,$^{3}$  
C.~Knigge,$^{4}$ 
J.~J.~Hermes,$^{5}$\newauthor
N.~P.~Gentile~Fusillo,$^{1,2}$
M.~A.~Hollands,$^{2}$
T.~Naylor,$^{6}$ 
I.~Pelisoli,$^{7}$
M.~R.~Schreiber,$^{8}$\newauthor 
S.~Toonen,$^{9}$
A.~Aungwerojwit,$^{10}$
E.~Cukanovaite,$^{2}$
E.~Dennihy,$^{11,12}$
C.~J.~Manser,$^{2}$\newauthor 
M.~L.~Pretorius,$^{13,14}$
S.~Scaringi,$^{15}$
O.~Toloza$^{2}$
\\
$^{1}$European Southern Observatory, Karl Schwarzschild Stra{\ss}e 2, Garching, 85748, Germany\\
$^{2}$Department of Physics, University of Warwick, Coventry, CV4 7AL, UK\\
$^{3}$Institute of Astronomy, University of Cambridge, Cambridge, CB3 0HA, UK\\
$^{4}$School of Physics and Astronomy, University of Southampton, Southampton, SO17 1BJ, UK\\
$^{5}$Department of Astronomy, Boston University, 725 Commonwealth Ave., Boston, MA 02215, USA\\
$^{6}$School of Physics, University of Exeter, Stocker Road, Exeter EX4 4QL, UK\\
$^{7}$Institut f\"ur Physik und Astronomie, Universit\"atsstandort Golm, Karl-Liebknecht-Str 24/25, D-14476 Potsdam, Germany\\
$^{8}$Instituto de F{\'i}sica y Astronom{\'i}a, Millennium Nucleus for Planet Formation (NPF), Universidad de Valpara{\'i}so, 2360102 Valparaiso, Chile\\
$^{9}$Institute for Gravitational Wave Astronomy, School of Physics and Astronomy, West office 237, Birmingham B15 2TT, Edgbaston, UK\\
$^{10}$Department of Physics, Faculty of Science, Naresuan University, Phitsanulok 65000, Thailand\\
$^{11}$University of North Carolina, Department of Physics and Astronomy, Chapel Hill, NC - 27599-3255, USA\\
$^{12}$Gemini Observatory, Southern Operations Center, Casilla 603, La Serena, Chile\\
$^{13}$Department of Astronomy, University of Cape Town, Private Bag X3, Rondebosch 7701, South Africa\\
$^{14}$South African Astronomical Observatory, PO Box 9, Observatory 7935, Cape Town, South Africa\\
$^{15}$Department of Physics and Astronomy, Texas Tech University, Lubbock, TX 79409-1051, USA}

\date{Accepted 2020 March 17. Received 2020 March 17; in original form 2019 July 26.}

\pubyear{2020}

\begin{document}
\label{firstpage}
\pagerange{\pageref{firstpage}--\pageref{lastpage}}
\maketitle

\begin{abstract}
We present the first volume-limited sample of cataclysmic variables (CVs), selected using the accurate parallaxes provided by the second data release (DR2) of the ESA \gaia space mission. The sample is composed of 42 CVs within 150~pc, including two new systems discovered using the \gaia data, and is $(77 \pm 10)\,$per cent complete. We use this sample to study the intrinsic properties of the Galactic CV population. In particular, the CV space density we derive, $\rho=(4.8^{+0.6}_{-0.8}) \times10^{-6}\,\pc$, is lower than predicted by most binary population synthesis studies.
We also find a low fraction of period bounce CVs, seven per cent, and an average white dwarf mass of $\langle M_\mathrm{WD} \rangle = (0.83 \pm 0.17)\,\mathrm{M}_\odot$. Both findings confirm previous results, ruling out the presence of observational biases affecting these measurements, as has been suggested in the past. 
The observed fraction of period bounce CVs falls well below theoretical predictions, by at least a factor of five, and remains one of the open problems in the current understanding of CV evolution.
Conversely, the average white dwarf mass supports the presence of additional mechanisms of angular momentum loss that have been accounted for in the latest evolutionary models.
The fraction of magnetic CVs in the 150\,pc sample is remarkably high at 36\,per cent. This is in striking contrast with the absence of magnetic white dwarfs in the detached population of CV progenitors, and underlines that the evolution of magnetic systems has to be included in the next generation of population models.
\end{abstract}




\section{Introduction}\label{sec:intro}
Cataclysmic variables (CVs) are compact interacting binaries containing a white dwarf accreting from a Roche-lobe filling donor star \citep[see][for a comprehensive review]{Warner1975}. In most systems, these companions are low-mass, late-type stars. If the white dwarf is not magnetised ($B \lesssim 1\,$MG), the mass lost from the donor forms an accretion disc around the white dwarf. In the presence of stronger magnetic fields, the disc is either truncated at the magnetospheric radius of the white dwarf ($1\,\mathrm{MG} \lesssim B \lesssim 10\,$MG) or is fully suppressed ($B \gtrsim 10\,$MG). In these magnetic CVs, known as intermediate polars (IPs) and polars, respectively, the accretion flow follows the field lines and accretes onto the white dwarf at its magnetic poles.

The evolution of CVs is, as for all types of interacting binaries, dictated by orbital angular momentum losses (AMLs) and by the internal structure of the companion star \citep{Rappaport+1983,Paczynski+1983, Spruit+1983,Knigge+2011}.
In fact, the time scale at which the secondary star loses mass is comparable to its thermal time scale, resulting in a donor which is slightly out of thermal equilibrium and hotter and bloated compared to an isolated main sequence star of the same mass. This deviation from thermal equilibrium is thought to be the cause of the major features observed in the CV orbital period distribution: the ``period gap'' and the ``period minimum'', as further detailed below.

As angular momentum is removed from the system, the orbital separation decreases and, consequently CVs evolve from long to short orbital periods \citep{Rappaport+1983,Paczynski+1983, Spruit+1983}.
At long orbital periods ($\porb \gtrsim 3\,$h) CV evolution is driven by magnetic wind braking (MB) and gravitational wave radiation (GWR). The ongoing mass transfer monotonously erodes the secondary star which, at $\porb \simeq 3\,$h, becomes fully convective. In the standard framework of CV evolution, it is assumed that a re-configuration of the magnetic fields on the donor results in a greatly reduced efficiency of MB from that point onwards, and the secondary star detaches from its Roche lobe. In the period range $2\,\mathrm{h} \lesssim \porb \lesssim 3\,$h, the so-called period gap, the system evolves as a detached binary whilst still losing angular momentum through GWR. Observational support for the disrupted MB hypothesis is provided by the observed properties of the post-common envelope binaries in the period range  $2\,\mathrm{h} \lesssim \porb \lesssim 3\,$h \citep{Schreiber+2010,Zorotovic+2016}.
At $\porb \simeq 2\,$h, the orbital separation is such that the donor fills its Roche lobe again and the accretion process resumes. 

Below the period gap, CVs keep evolving towards shorter orbital periods until they reach the period minimum, $\porb \simeq 80\,$min. At this stage, the time-scale on which the secondary star loses mass becomes much shorter compared to its thermal time-scale. The donor is driven out of thermal equilibrium and stops shrinking in response to the mass loss. Consequently the system starts evolving back towards longer orbital periods, becoming a ``period bouncer''.  

With MB being much more efficient than GWR in removing angular momentum from the system, CVs above the period gap are predicted to have mass accretion rate orders of magnitudes higher ($\dot{M}\,\sim\,10^{-9} - 10^{-8} \,\mathrm{M}_\odot \, \mathrm{yr}^{-1}$, \citealt{Spruit+1983}) than those of the CVs below the period gap ($\dot{M} \sim 5 \times 10^{-11} \,\mathrm{M}_\odot \, \mathrm{yr}^{-1}$, \citealt{Patterson1984}). While this is roughly in agreement 
with the accretion rates estimated from observations \citep{Townsley+2009,Pala+2017}, the theoretical framework outlined above fails to reproduce a number of observational properties of the Galactic population of CVs: (i) the predicted fractions of CVs above and below the period gap ($\simeq 1\,$per cent and $\simeq 99\,$per cent, respectively, \citealt{deKool1992,Kolb1993,Howell+2001}) are in clear disagreement with the observations (e.g. $\simeq 23\,$per cent and $\simeq 77\,$per cent, \citealt{Gaensicke+2009}, though the observed samples are typically magnitude-limited, and hence biased towards more luminous CVs); (ii) period bouncers are expected to be the main component ($\simeq 40-70\,$per cent) of the present-day Galactic CV population but only a small number of such systems has been identified \citep{pattersonetal05-2,Unda-Sanzana2008, Littlefair+2006, Patterson2011, Kato+2015, Kato+2016, McAllister+2017, Neustroev+2017, Pala+2018}; (iii) there are clues of the presence of additional AML mechanisms that are not accounted for by the standard model of CV evolution \citep{Patterson1998,Knigge+2011,Schreiber+2016,Pala+2017,Zorotovic+2017,Belloni+2018,Liu+2019}, although the physical origin of this enhanced AML is still unclear.

A key parameter that provides stringent constraints on the models of CV formation and evolution is their space density, $\rho_0$. Binary population synthesis studies carried out by \citet{deKool1992} and \citet{Politano1996} suggested CV space densities of $\simeq 2\times10^{-5} - 2.0\times10^{-4}\,\pc$. More recent works by \citet{Goliasch-Nelson2015}, which also accounts for the presence of CVs containing nuclear evolved donors, and by \citet{Belloni+2018} provide an estimate of $\simeq (1.0\pm0.5)\times10^{-5}\,\pc$ and $\lesssim 2 \times10^{-5}\,\pc$, respectively, comparable to the earlier results.

These predicted values are systematically larger than the ones derived from observations. For example, \citet{Thomas+1998} used the \textit{ROSAT} All Sky Survey to infer $\rho_0\simeq6.1\times10^{-7}\,\pc$ for polars. Later studies based on the \textit{ROSAT} Bright Survey (RBS) and the \textit{ROSAT} North Ecliptic Pole (NEP) suggested $\rho_0=4{+6\atop-2}\times 10^{-6}\pc$ \citep{Pretorius+2007a,Pretorius+2012}. Most recently,  \citet{Hernandez+2018} estimated an upper limit on the space density of period bounce CVs from a search for eclipsing systems in Stripe\,82 from the Sloan Digital Sky Survey (SDSS, \citealt{York+2000}), finding $\rho_0\la2\times10^{-5}\,\pc$. 

\begin{figure*}
\includegraphics[width=\textwidth]{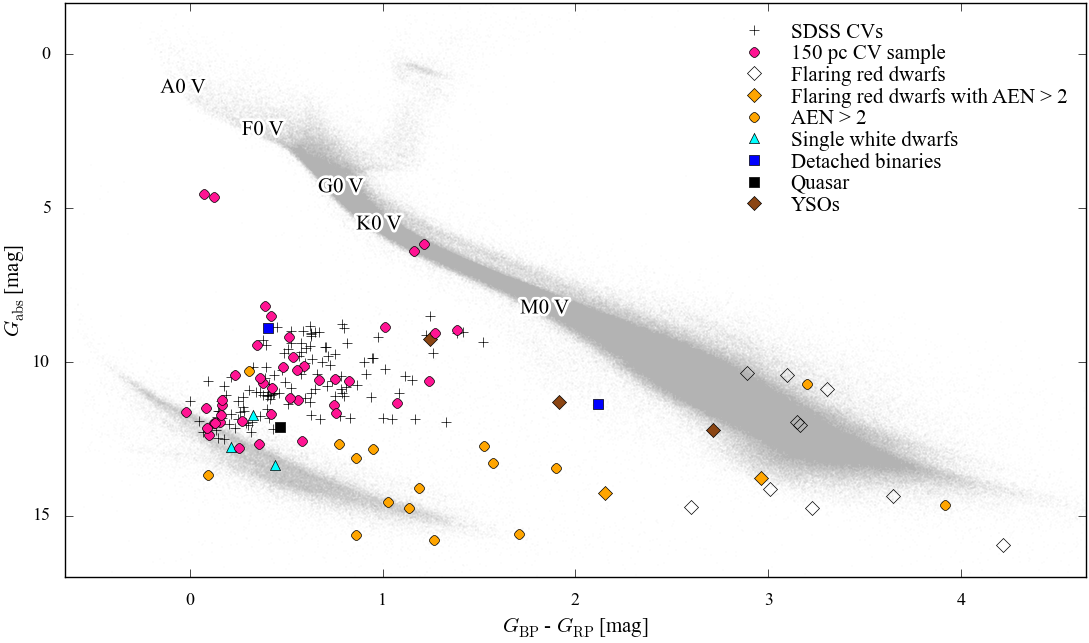}
\caption{Hertzsprung-Russell diagram of the \gaia sources with reliable astrometry (Equation~\ref{eq:clean}) located within 150\,pc (grey), showing the position of the 150~pc CVs (pink dots, Section~\ref{sec:150pc}) and of the different types of contaminants (Section~\ref{sec:sample}) that  have been mistakenly identified as CVs or CV candidates within the literature (Appendices~\ref{ap:frd}, \ref{ap:yso} and \ref{ap:ncvs}). For comparison, SDSS CVs with accurate and clean SDSS photometry (as defined in Section~\ref{subsec:completeness}) are shown by the black crosses, however the majority of these systems are at distances $d > 150\,$pc. Systems discarded for having high astrometric excess noise (AEN) in \gaia are shown in yellow. 
}\label{fig:hr_all}
\end{figure*}

In the past, an accurate measurement of the CV space density has been challenged by the lack of accurate distances. In April 2018, the European Space Agency (ESA) \gaia space mission has delivered parallaxes for more than one billion stars in its second data release DR2 \citep{Prusti+2016,Brown+2018}, providing the first opportunity to construct a volume-limited sample of CVs and to derive their intrinsic properties. \citet{Schwope2018} carried out the first application of the \gaia data in this context, and, using the distances from \gaia for a X-ray selected sample of CVs from RBS, NEP and \textit{Swift}/BAT survey, derived the space densities of IPs, $\rho_0 < 1.3\times 10^{-7}\pc$, and non-magnetic CVs, $\rho_0 < 5.1\times 10^{-6}\pc$. These results are based on the assumptions that the X-ray selected sample is complete and representative of the intrinsic population. However, as discussed by \citet{Pretorius+2012}, it is possible that a large fraction of faint CVs may not have been detected in the RBS and NEP surveys, and that the space density derived from the corresponding X-ray selected CV sample could be easily underestimated by a factor $\simeq 2$.

Here we present a study of the first volume-limited sample of CVs within 150\,pc, selected by combining the \gaia DR2 parallaxes and the available information from a large number of spectroscopic and photometric surveys.

The sample contains a total of 42 objects and provides the first direct insight into the intrinsic properties of the Galactic population of CVs.

\begin{figure*}
\includegraphics[width=\textwidth]{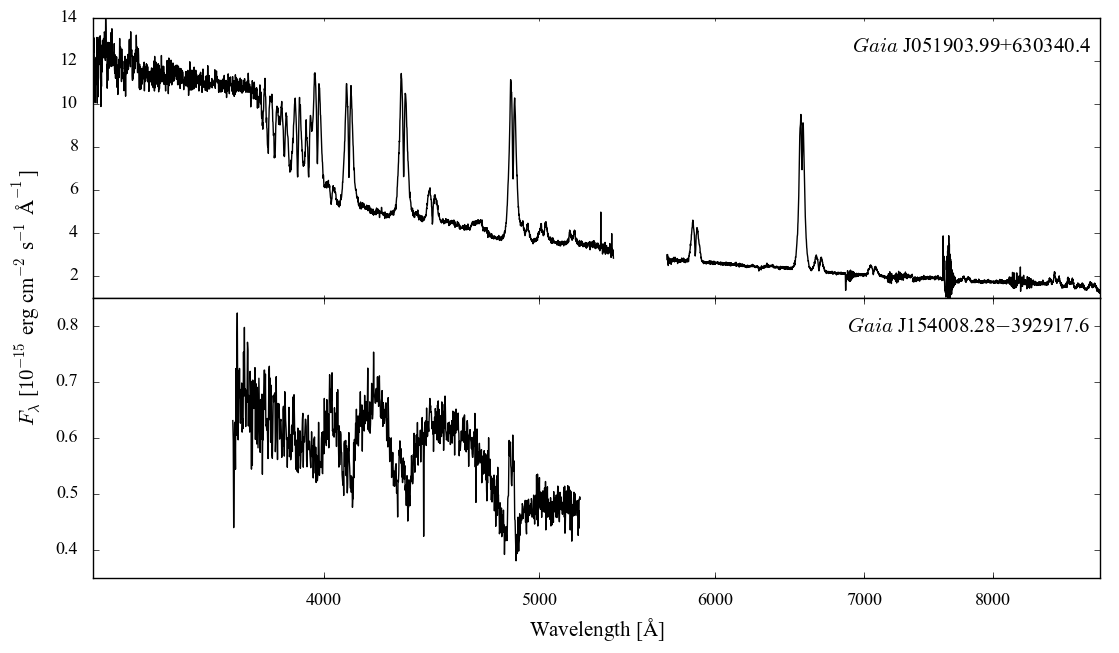}
\caption{Identification spectra of the two new CVs we discovered within 150~pc. \gaia\,J051903.99+630340.4 (top) presents the typical spectral appearance of an SU\,UMa CV as it is also confirmed by its orbital period, $P_\mathrm{orb} \simeq 126\,$min (Appendix~\ref{ap:ncvs}), which locates it below the period gap. 
Dwarf nova outbursts of this system have been detected by ASAS-SN in February 2012, March 2016 and September 2018, during which the system brightened, on average, by $\simeq 5.9\,$mag. However, in the ASAS-SN observations, the CV is blended with a bright ($V = 10.1\,$mag) nearby companion (within $\simeq 7\arcsec$, see Section~\ref{sec:triplets}). Consequently, during each dwarf nova outburst, only a variation of $\simeq 0.6\,$mag is observed in the overall brightness recorded by ASAS-SN of the two stars blended together, thus preventing an earlier detection of \gaia\,J051903.99+630340.4 as a CV.
Conversely, \gaia\,J154008.28--392917.6 (bottom) is a low accretion rate system of the WZ\,Sge type, likely located at the period minimum (note its similarity with e.g. EZ\,Lyn in Figure~\ref{fig:sdss_spectra}), which has never been observed in outburst so far.}\label{fig:new_gaia_cvs}
\end{figure*}

\section{Sample selection}\label{sec:sample}
The \gaia space mission provides the first opportunity to construct a volume-limited sample of CVs that allows us to infer the intrinsic properties of the Galactic CV population. 

The first step towards this goal is the choice of an optimal volume, which is sufficiently large to be representative of the entire CV population and robust against small number statistics, and not subject to distance (magnitude) related observational biases. 

The volume enclosed within 150\,pc represents a good compromise between the two requirements: assuming the typical space density derived from binary population synthesis studies, $\lesssim 2 \times10^{-5}\,\pc$ \citep{Belloni+2018}, it is expected to contain $\simeq 140$ CVs, and at $d=150$\,pc even the systems with the lowest accretion rates\footnote{e.g. QZ\,Lib, one of the few period bouncers known, is located at a distance of $187\,$pc ($\varpi = 5.3 \pm 0.3\,$mas) and is as bright as $G = 18.9\,$mag \citep{Pala+2018}.} should be bright enough to have accurate astrometric solutions (typical uncertainties on the parallaxes are $\simeq 0.2\,$mas for $G \lesssim 19\,$mag, see Figure~7 in \citealt{Brown+2018}). Restricting $d\le150$\,pc also reduces the uncertainties in the derived space densities related to the unknown age and scale height of the CV population (see Section~\ref{sec:150pc}).  

The next step is to compile a list of all CVs and CV candidates that could plausibly be within $d\le150$\,pc. CVs are mainly discovered thanks to their outbursting properties. In fact, CV accretion discs undergo thermal instabilities called dwarf nova outbursts \citep{Osaki1974,Meyer+1984,Hameury+1998},
during which CV systems brighten up to 2--9 mag and these outbursts can last for days up to weeks \citep{Warner1975,Maza+1983}. 
Many surveys search the sky nightly for transient events, such as the Catalina Real-time Transient Survey (CRTS, \citealt{Drake+2009}), the All-Sky Automated Survey and the All-Sky Automated Survey for Supernovae (ASAS and ASAS-SN, \citealt{Pojmanski1997,Shappee+2014,Kochanek+2017}), the Mobile Astronomical System of TElescope Robots (MASTER, \citealt{Lipunov+2010}), the Palomar Transient Factory (PTF, \citealt{Law+2009}) and the Intermediate PTF (iPTF, \citealt{Kulkarni2013}), the Asteroid Terrestrial-impact Last Alert System (ATLAS; \citealt{Tonry+2018}), the Zwicky Transient Facility (ZTF; \citealt{Graham+2019}) and the Gaia Photometric Science Alerts \citep{Hodgkin+2013}, which have been very successful in the identification of thousands of CVs in outbursts \citep[e.g.][]{Breedt+2014}. 

Alternatively, CVs can be identified thanks to their blue colours \citep[see e.g.][]{Boris2005} and their X-ray emission (the latter favouring magnetic systems since the X-ray emission arises mainly from small, hot region near the magnetic poles of polars and IPs).
Finally, CVs are one of by-products of SDSS, with over 300 new CVs discovered in the last decade via the detection of strong emission disc lines in their spectra \citep{Szkody+2002, Szkody+2003, Szkody+2004, Szkody+2005, Szkody+2006, Szkody+2007, Szkody+2009, Szkody+2011}. Whilst the spectroscopic method is less affected by selection biases, it is the most expensive in term of telescope time.

\begin{figure*}
\includegraphics[width=\textwidth]{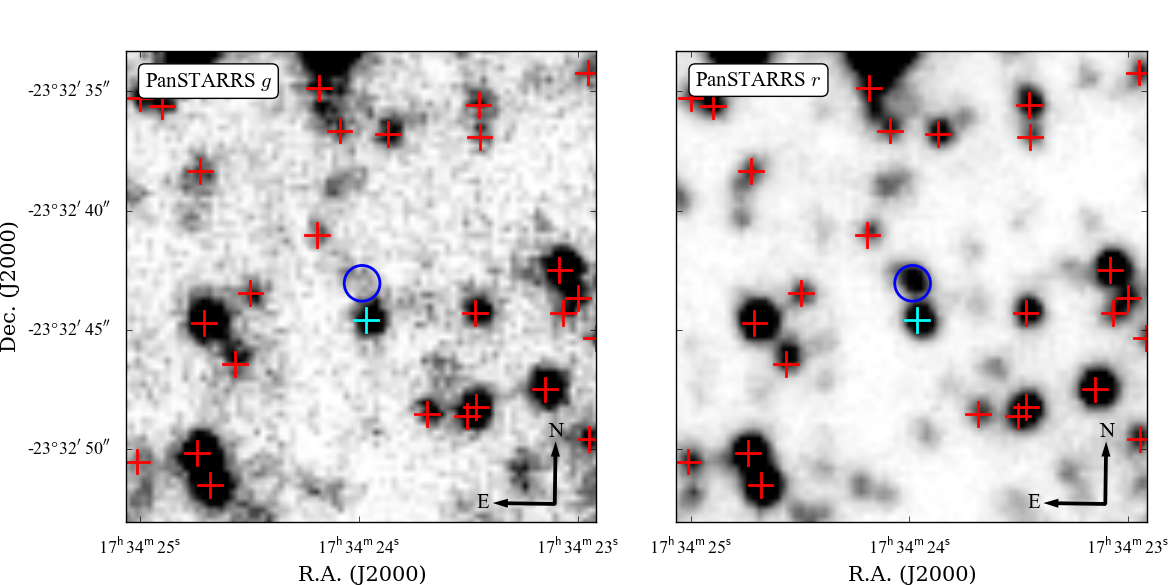}
\caption{PanSTARRS $g$-band (left) and $r$-band (right) images of OGLE-BLG-DN-0040. 
The position of OGLE-BLG-DN-0040 is highlighted with a blue circle, while the crosses show the \gaia detections, one of which (cyan cross) is located only $\simeq 1.4\arcsec$ away from OGLE-BLG-DN-0040. However, OGLE-BLG-DN-0040 is fainter than the \gaia detection limit in quiescence so does not have an entry in the \gaia DR2 archive. This is actually a spurious association, as can be clearly seen in the $r$-band image.}\label{fig:ogle0040}
\end{figure*}

New systems are continuously discovered and an up-to-date catalogue of these discoveries is missing.
We therefore searched the catalogues of the aforementioned transient surveys and the International Variable Star Index (VSX, \citealt{Watson+2006}, compiled by the American Association of Variable Star Observers, AAVSO) to collect all information regarding newly identified CVs and CV candidates. Combining these findings with the Ritter \& Kolb catalogue \citep{RitterKolb}, a collection of the observational properties of all CVs with an orbital period determination (1429 systems), we built a list of $\simeq 8000$ systems. 

We cross-matched this list with the \gaia DR2 catalogue. In order to take into account proper motions and the low precision ($\simeq 10\arcsec$) of the coordinates reported in some of the catalogues, we first performed a cross-match with a 30\arcsec\, search radius, resulting in $\simeq 364\,000$ objects. Using the \gaia proper motions and radial velocities (whenever available), we calculated the corresponding coordinates at epoch 2000. We then performed a second cross-match between our list of CVs and this catalogue, selecting the closest source within 10\arcsec\, radius, which resulted in $\simeq 6400$ objects. Since our focus is on the properties of the 150~pc CV sample, we applied a cut in parallax: 
\begin{equation}\label{eq:parallax_cut}
\varpi + 3\sigma_\varpi >= 6.66\,\mathrm{mas}
\end{equation} 
This selection returned 166 systems which, within their parallax uncertainties, are located within 150~pc (Figure~\ref{fig:hr_all}). 

More than half of these 166 systems have been identified as CV candidates in transient surveys  because they have shown one or more brightenings that resembled a dwarf nova outburst, but have no photometric or spectroscopic follow-up confirming their CV nature. Therefore, for each system, we inspected the literature, the CRTS, AAVSO and ASAS-SN light curve archives for their outburst history, and SDSS for serendipitous spectroscopy. In this way, we identified 28 objects that are mistakenly classified as CVs or CV candidates in the literature (Table~\ref{tab:discarded}), many of them single low-mass stars which show flaring phenomena that can be mistaken for dwarf nova outbursts. More details are provided in the Appendices~\ref{ap:frd}, \ref{ap:yso} \& \ref{ap:ncvs}.

Finally, we searched for \gaia sources in the white dwarf locus of the HR diagram (defined by the colour cuts from \citealt{Mark+2018}, see their equations 2, 3 and 4) that showed large anomalous uncertainties in their $G$ magnitudes, weighted over the number of observations in the given band.  Such large values can be interpreted as intrinsic stellar variability arising e.g. from pulsations, debris transits, magnetism or on-going accretion (Hermes et al., in preparation). Among the different candidates, two systems stood out because of their large intrinsic variability and for being overluminous to the canonical white dwarf cooling tracks. 
Spectroscopic follow-up later confirmed these as new CVs located within 150~pc: \gaia\,J051903.99+630340.4 and \gaia\,J154008.28--392917.6 (Figure~\ref{fig:new_gaia_cvs} and Appendix~\ref{ap:new_cvs}, which we named with the identifier \gaia~JHHMMSS.SS$\pm$DDMMSS.S, defined with respect to their coordinates at equinox 2000 and epoch 2000, as computed using the \gaia ICRS coordinates and proper motions).

\subsection{Astrometric accuracy}\label{sec:aen}
Along with positions, magnitudes and proper motions, i.e. the kinematic parameters used to obtain the astrometric solution for each source, \gaia DR2 provides a series of additional parameters that allow to evaluate the accuracy of this solution.
In particular, the \texttt{astrometric\_excess\_noise} (AEN) represents the errors introduced in the astrometric modelling \citep[see][]{Lindegren+2012} and, ideally, should be zero. A possible selection to remove sources with poor astrometric solution is to impose \texttt{astrometric\_excess\_noise < 1} \citep{Lindegren+2018}. However, we noticed that Z\,Cha, for which the \gaia parallax ($\varpi = 8.7 \pm 0.1\,$mas) implies a distance of $116 \pm 2\,$pc, in good agreement with the distance estimated by \citet{Beuermann2006}, $112 \pm 8\,$pc, has \texttt{astrometric\_excess\_noise = 1.08}. Therefore, in order to build a sample that is the most complete as possible we applied a more generous cut, considering the parallaxes of those sources for which \texttt{astrometric\_excess\_noise > 2} as unreliable.

\subsection{Spurious \gaia detections}\label{sec:spurious}
Ten CV candidates are located in crowded regions and the cross-match between our sample and the \gaia catalogue returned a spurious detection (Figure~\ref{fig:ogle0040}). These sources are actually fainter than the \gaia detection limit and do not have an entry in the \gaia DR2 archive. 
Our list also contained EU\,Cnc for which \gaia DR2 provides a parallax of $\varpi = 1.8 \pm 2.5\,$mas and therefore satisfies the condition in Equation~\ref{eq:parallax_cut}. However, this CV is located in an open cluster for which the distance has been determined as $d = 785 \pm 50\,$pc \citep{Belloni+1993} and hence we discarded this system. 

\begin{table*}
\caption{150~pc CV sample from \gaia DR2, sorted by increasing distance. The CV types are as follows: UG, dwarf nova of U\,Gem sub-type; UGSU, dwarf nova of SU\,UMa sub-type; UGWZ, dwarf nova of WZ\,Sge sub-type; IP, intermediate-polar; AM, polar; NL, novalike; AM\,CVn, AM\,CVn star. SDSS $ugriz$ photometry of the systems highlighted with a star is used to assess the completeness of the \gaia 150~pc sample (see Section~\ref{sec:gaia_sdss}).}\label{tab:cv_table}
\begin{tabular}{@{}llccccccccc@{}}
 \toprule
 System         & \multicolumn{1}{c}{\gaia DR2 ID} & $\varpi$ & $\sigma_\varpi$ & $P_\mathrm{orb}$ &   Type &  $G$ & $G_\mathrm{BP}$ & $G_\mathrm{RP}$ & Distance \\
              &  &  (mas) &  (mas) &  (min) &    &  (mag) & (mag) &  (mag) & (pc) \\
\midrule 
WZ\,Sge & 1809844934461976832 & 22.16 &  0.04 & 81.63 & UGWZ & 15.21 & 15.21 & 15.06 & $45.13 \pm  0.08$ \\
VW\,Hyi & 4653893040002306432 & 18.53 &  0.02 & 106.95 & UGSU & 13.84 & 13.94 & 13.45 & $53.96 \pm  0.06$ \\
EX\,Hya & 6185040879503491584 & 17.56 &  0.04 & 98.26 & IP & 13.21 & 13.23 & 12.88 & $ 56.9 \pm   0.1$ \\
GP\,Com & 3938156295111047680 & 13.73 &  0.06 & 46.57 & AM\,CVn & 15.95 & 15.89 & 15.91 & $ 72.8 \pm   0.3$ \\
V455\,And & 1920126431748251776 & 13.24 &  0.06 & 81.08 & IP & 16.06 & 16.13 & 15.71 & $ 75.5 \pm   0.3$ \\
GD\,552 & 2208124536065383424 & 12.35 &  0.05 & 102.73 & UGWZ & 16.46 & 16.46 & 16.18 & $ 81.0 \pm   0.3$ \\
ASASSN-14dx* & 2488974302977323008 & 12.34 &  0.04 & 82.81 & UGWZ & 14.96 & 14.92 & 14.69 & $ 81.0 \pm   0.3$ \\
AM\,Her & 2123837555230207744 & 11.40 &  0.02 & 185.65 & AM & 13.58 & 13.86 & 12.85 & $ 87.8 \pm   0.1$ \\
IX\,Vel & 5515820034889610112 & 11.04 &  0.03 & 279.25 & NL &  9.32 &  9.34 &  9.27 & $ 90.6 \pm   0.2$ \\
OY\,Car & 5242787486412627072 & 11.01 &  0.03 & 90.89 & UGSU & 15.62 & 15.64 & 15.21 & $ 90.8 \pm   0.2$ \\
AE\,Aqr & 4226332451596335616 & 10.97 &  0.06 & 592.78 & IP & 10.95 & 11.47 & 10.26 & $ 91.2 \pm   0.5$ \\
U\,Gem & 674214551557961984 & 10.71 &  0.03 & 254.74 & UG & 13.91 & 14.38 & 13.11 & $ 93.4 \pm   0.3$ \\
V396\,Hya & 3503987633230546688 & 10.69 &  0.15 & 65.10 & AM\,CVn & 17.66 & 17.70 & 17.45 & $   94 \pm     1$ \\
BW\,Scl & 2307289214897332480  & 10.60 &  0.10 & 78.23 & UGWZ & 16.26 & 16.26 & 16.10 & $ 94.4 \pm   0.9$ \\
V627\,Peg & 1800384942558699008 & 10.03 &  0.07 & 78.51 & UGWZ & 15.67 & 15.65 & 15.27 & $ 99.7 \pm   0.7$ \\
AR\,UMa & 783921244796958208 &  9.87 &  0.12 & 115.92 & AM & 16.26 & 16.35 & 15.78 & $  101 \pm     1$ \\
1RXS\,J105010.3--140431 & 3750072904055666176 &  9.14 &  0.11 & 88.56 & UGWZ & 17.17 & 17.21 & 17.08 & $  109 \pm     1$ \\
TCP\,J21040470+4631129 & 2163612727665972096 & 9.13 & 0.12 & $77.07^{\mathrm{(a)}}$ & UGWZ & 17.77 & 17.87 & 17.29 & $109 \pm 2$ \\
V2051\,Oph & 4111991385628196224 &  8.90 &  0.07 & 89.90 & UGSU & 15.37 & 15.46 & 14.87 & $112.4 \pm   0.9$ \\
V834\,Cen & 6096905573613586944 &  8.90 &  0.21 & 101.52 & AM & 16.66 & 16.82 & 16.07 & $  113 \pm     3$ \\
GW\,Lib & 6226943645600487552 &  8.87 &  0.08 & 76.78 & UGWZ & 16.49 & 16.49 & 16.32 & $  113 \pm     1$ \\
ST\,LMi & 3996419759863758592 &  8.83 &  0.08 & 113.89 & AM & 16.13 &   -- &   -- & $  113 \pm     1$ \\
SS\,Cyg & 1972957892448494592 &  8.72 &  0.05 & 396.19 & UG & 11.69 & 12.11 & 10.95 & $114.6 \pm   0.6$ \\
V884\,Her & 4503256687122329088 &  8.69 &  0.02 & 113.01 & AM & 13.49 & 13.57 & 13.18 & $115.1 \pm   0.3$ \\
Z\,Cha & 5210507882302442368 &  8.66 &  0.12 & 107.28 & UGSU & 15.85 & 15.94 & 15.19 & $  116 \pm     2$ \\
\gaia\,J051903.99+630340.4 & 285957277597658240 &  8.59 &  0.04 & 126: & UGSU & 15.17 & 15.30 & 14.77 & $  116.4 \pm 0.5 $ \\
V2301\,Oph & 4476137370261520000 &  8.24 &  0.08 & 112.97 & AM & 16.75 & 16.94 & 15.86 & $  121 \pm     1$ \\
V893\,Sco & 6039131391540808832 &  8.06 &  0.05 & 109.38 & UGSU & 14.65 & 14.76 & 14.25 & $124.1 \pm   0.8$ \\
QZ\,Vir* & 3800596876396315648 &  7.81 &  0.07 & 84.70 & UGSU & 16.06 & 16.12 & 15.76 & $  128 \pm     1$ \\
V1040\,Cen & 5343601913741261312 &  7.80 &  0.03 & 87.11 & UGSU & 14.04 & 14.11 & 13.69 & $128.1 \pm   0.4$ \\
SDSS\,J125044.42+154957.3 & 3934459045528378368 & 7.79 & 0.18 & 86.3 & AM & 18.22 & 18.30 & 17.95 & $129 \pm 3$  \\
V379\,Tel & 6658737220627065984 &  7.65 &  0.07 & 101.03 & AM & 16.19 & 16.69 & 15.45 & $  131 \pm     1$ \\
BL\,Hyi & 4697621824327141248 &  7.65 &  0.07 & 113.64 & AM & 17.25 & 17.45 & 16.70 & $  131 \pm     1$ \\
MR\,Ser* & 1203639265875666304 &  7.59 &  0.05 & 113.47 & AM & 16.23 & 16.47 & 15.64 & $131.8 \pm   0.8$ \\
V3885\,Sgr & 6688624794231054976 &  7.54 &  0.08 & 298.31 & NL & 10.25 & 10.28 & 10.16 & $  133 \pm     1$ \\
\gaia\,J154008.28--392917.6 & 6008982469163902464 &  7.49 &  0.11 &   -- & UGWZ & 17.36 & 17.39 & 17.23 & $  134 \pm     2$ \\
VV\,Pup & 5719598950133755392 &  7.30 &  0.05 & 100.44 & AM & 15.93 & 16.04 & 15.48 & $137.0 \pm   0.9$ \\
VY\,Aqr & 6896767366186700416 &  7.24 &  0.14 & 90.85 & UGSU & 16.86 & 16.96 & 16.44 & $  138 \pm     3$ \\
IP\,Peg & 2824150286583562496 &  7.08 &  0.05 & 227.82 & UG & 14.71 & 15.27 & 13.88 & $141.2 \pm   1.0$ \\
HT\,Cas & 426306363477869696 &  7.07 &  0.06 & 106.05 & UGSU & 16.35 & 16.48 & 15.81 & $  141 \pm     1$ \\
SDSS\,J102905.21+485515.2 & 834947865750806272 &  6.96 &  0.24 & 91.33 & UGWZ & 18.16 & 17.94 & 17.84 & $  144 \pm     5$ \\
EZ\,Lyn* & 935056333580267392 &  6.87 &  0.15 & 84.97 & UGWZ & 17.81 & 17.84 & 17.71 & $  146 \pm     3$ \\
V379\,Vir* & 3699606286708406912 & 6.7 & 0.2 & 88.4 & AM & 18.01 & 18.02 & 17.93 & $150 \pm 4$\\
V355\,UMa & 1558322303741820928 & 6.66 & 0.09 & 82.52 & UGWZ & 17.38 & 17.35 & 17.27 & $150 \pm 2$ \\
\bottomrule
\end{tabular}
\begin{tablenotes}
\item \textbf{Notes.} (a): spectroscopic period (ATel \#13009).
\end{tablenotes}
\end{table*}

\begin{figure}
\includegraphics[trim={3.5cm 0.5cm 0 1cm},clip,width=0.56\textwidth]{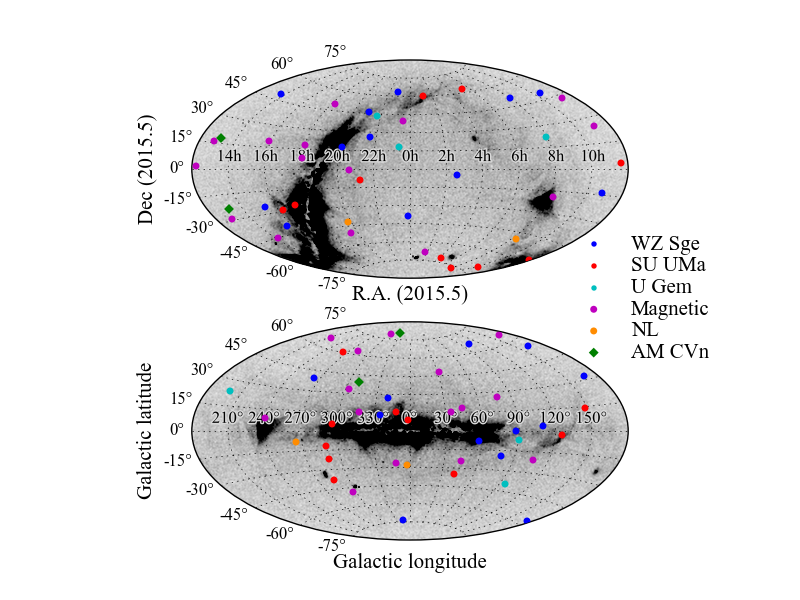}
\caption{The spatial distribution of CVs in the 150\,pc sample in equatorial (top) and Galactic (bottom) coordinates shows no evident correlation with the different sub-types, suggesting that the 150~pc CV sample is not affected by obvious selection biases.}\label{fig:ra_dec}
\end{figure}

\section{The 150~pc CV sample}\label{sec:150pc}
\gaia DR2 provided parallaxes for about 1.3 billion sources. Converting these measurements into distances is not always trivial as the mere inversion of the parallax can introduce some biases in the distance estimate, especially when the fractional error on the parallax is larger than 20~per cent \citep[see e.g.][]{Bailer-Jones2015,Luri+2018}. This is the case for many systems in our sample, in particular for those that are further away and have poor parallax measurements. Therefore, to estimate the distances, we used a statistical approach in which, following the prescription by \citet{Bailer-Jones2015}, we defined a likelihood and an exponentially decreasing volume density prior. The latter contains the Galactic CV scale height, $h$, which is a function of the system age: older CVs (i.e. period bouncers and short period systems that have not evolved through the period minimum yet, $\porb \lesssim 2\,$h) have larger scale heights, $h \simeq 260-450\,$pc, while younger CVs (i.e. long period systems, $\porb \gtrsim 2\,$h) have smaller scale heights, $h \simeq 120\,$pc \citep{Pretorius+2007b}. For the systems with a measured orbital period we defined the relative scale heights following \citet{Pretorius+2007b}:
\begin{equation}\label{eq:scale_heights}
h = \begin{cases} 120~\mathrm{pc} & \mbox{for long period systems (} \porb \gtrsim 2\,\mbox{h)} \\ 
                  260~\mathrm{pc} & \mbox{for short period systems (}\porb \lesssim 2\,\mbox{h)} \\
                  450~\mathrm{pc} & \mbox{for period bouncer CVs} \end{cases}
\end{equation}
Our sample contains three period bouncer candidates, GD\,552 \citep{Unda-Sanzana2008},  SDSS\,J102905.21+485515.2 \citep{Thorstensen+2016} and 1RXS\,J105010.3--140431 \citep{Patterson2011,Pala+2017}, for which we assumed a scale height of $h = 450\,$pc. An orbital period determination is not available for \gaia\,J154008.28--392917.6. Nonetheless, its spectral appearances suggest that this is a low accretion rate systems of the WZ\,Sge sub-type. These CVs are found below the period gap ($\porb \lesssim 2\,$h) and thus, for \gaia\,J154008.28--392917.6, we assumed a scale height of $h = 260\,$pc. 

We determined the distance as the median and the corresponding uncertainties as the 16th and the 84th percentiles of the posterior distribution of each system. From the posterior distribution, we also determined the probability for each object to be located within 150\,pc.
According to these distances, we discarded 55 systems (Tables~\ref{tab:cv_excluded}) for which $d > 150\,$pc. 

The final 150~pc sample (Table~\ref{tab:cv_table}) consists of 42 CVs and two AM\,CVn systems\footnote{An AM\,CVn system consists of a white dwarf accreting from another white dwarf or a partially degenerate helium star, often via an accretion disc. Their orbital periods are shorter than that of CVs and their optical spectra do not contain any hydrogen.} (GP\,Com and V396\,Hya). Although there are many similarities between AM\,CVn stars and CVs, the formation channels and evolutionary histories of these two classes of systems are different and we do not include the two AM\,CVn in the following discussion (see \citealt{Ramsay+2018} for a detailed study of these stars using the \gaia DR2 data).

Figure~\ref{fig:ra_dec} show the spatial distribution of this sample and it is colour-coded according to the CV type. 
The systems appear to be uniformly distributed on the sky, with no evident correlation between their location and their sub-types. In order to evaluate whether the position of the 150\,pc CVs on the sky is consistent with a uniform distribution from a statistical point of view, we generated 1000 samples, each consisting of $10^5$ objects uniformly distributed in a volume enclosed within 150\,pc, assuming an exponentially decreasing space density (Equation~\ref{eq:gal_model}) with $h= 280\,$pc, i.e. the average value of the scale heights of different CV subtypes (Equation~\ref{eq:scale_heights}). For each sample, we randomly selected 42 objects and applied a two sample Kolmogorov-Smirnov (KS) test between the pairwise cumulative distributions of these objects and that of the CVs in the 150\,pc sample. We found that the average KS statistics is $D \simeq 0.03$, lower than the 95\,per cent critical level, $D_\mathrm{critic} \simeq 0.29 $, with a \textit{p-value $\simeq 0.7$}. This implies that the null hypothesis, i.e. uniform distribution on the sky, cannot be rejected thus suggesting that the 150~pc sample is not affected by obvious selection biases.

\section{Completeness}\label{subsec:completeness}
A key property of any sample used to determine a space density is its \textit{completeness}, in combination with a good understanding of potential observational selection effects. We use two independent tests to assess the completeness of the 150~pc CV sample. 

\subsection{Previously known CVs with distances}
A simple assessment of the completeness of the 150\,pc CV sample is to establish the fraction of known CVs with \textit{reliable} pre-\gaia distance measurements of $d\le150$\,pc recovered by \gaia. Except parallaxes, all distance measurements are indirect, and we briefly review the main methods used for CVs to qualify what we consider as reliable distances\footnote{\citet{Ak+2007} proposed a period-luminosity-colour relation, i.e. an alternative empirical calibrations in which the absolute magnitude of the system is determined from its period and its 2MASS (Two Micron All Sky Survey, \citealt{Skrutskie+2006}) colours. However, its application to short-period dwarf novae appeared to be problematic \citep{Patterson2011} and we excluded the distances derived with this method from the following analysis.}.

\begin{figure}
\includegraphics[width=0.45\textwidth]{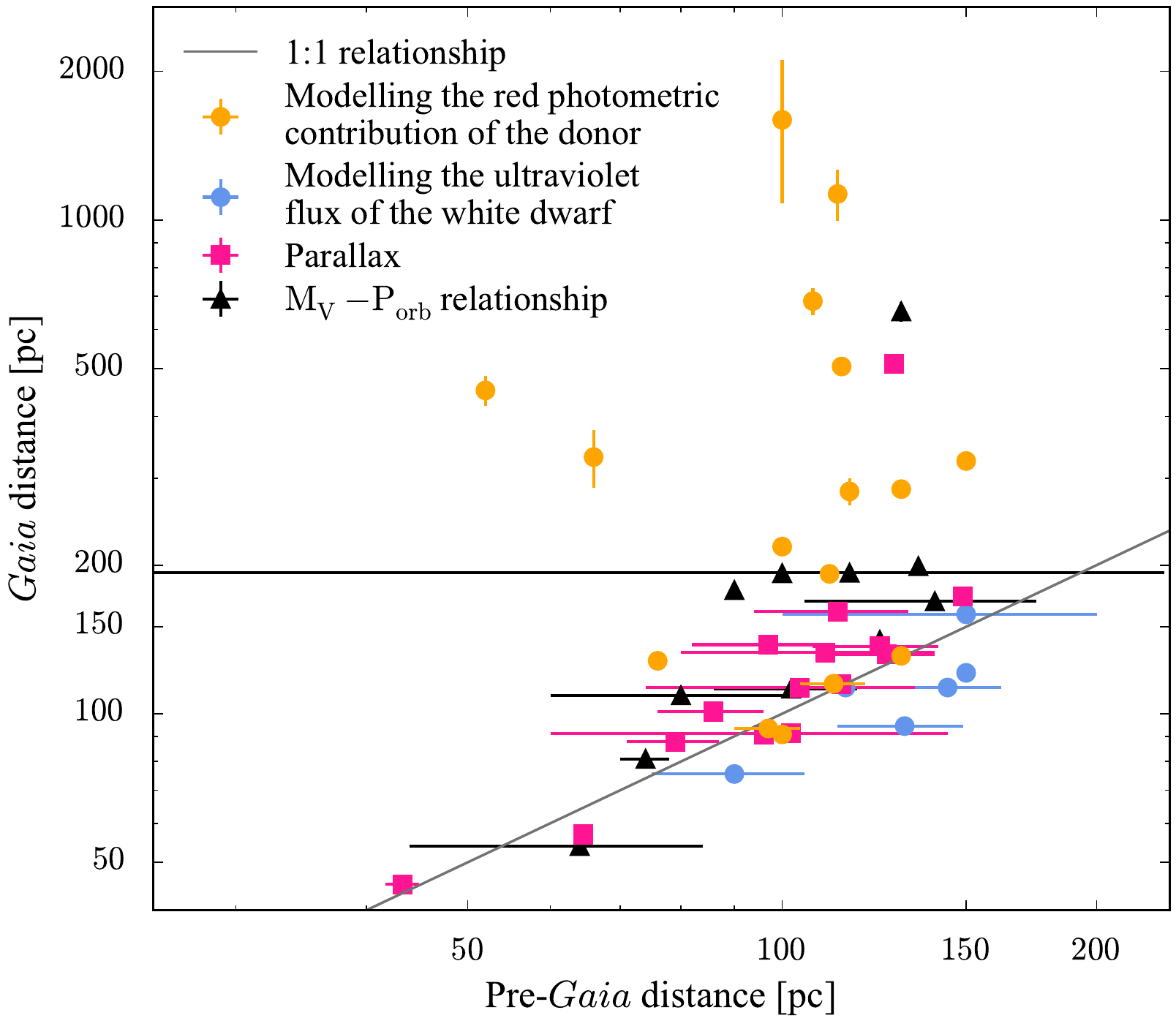}
\caption{Comparison between pre-\gaia distances estimated using different methods and distances based on the \gaia parallaxes. The most accurate pre-\gaia distances have been determined from ground- and space-based parallax measurements (pink). Whereas modelling ultraviolet spectra of white dwarf dominated CVs (blue) also provides reliable distance measurements, estimates based on modelling the red/infrared photometric contribution of the donor stars (orange) systematically overestimate the distance.  Although subject to large uncertainties, the distance estimates from the $M_V - P_\mathrm{orb}$ relationship are consistent with the \gaia determinations for most of the systems in the sample.}\label{fig:pre-post-gaia}
\end{figure}

\begin{enumerate}[a)]
\item Accurate trigonometric parallaxes from both space- \citep{Duerbeck1999,Harrison2004} and ground-based observations \citep{Thorstensen2003,Thorstensen+2008}. 

\item Modelling the ultraviolet flux of the white dwarf. This method requires a clean detection of the white dwarf which, owing to the strong contamination from the accretion disc in the optical, is best detected in the ultraviolet. By performing a fit to the spectroscopic data, the distance to the system can be measured by assuming a mass-radius relationship, since the scaling factor between the best-fitting model is related to the ratio between the distance and the stellar radius \citep[e.g.][]{Boris+1999a}.

\item Modelling the red/infrared photometric contribution of the donor.  The emitting area of the secondary is constrained by the Roche geometry, i.e. by the orbital period and the mass ratio. When the secondary is detected in the near-infrared, the distance to the system can be estimated using the ``Bailey method'' \citep{Bailey1981} from its $K$ magnitude in combination with the mass-radius relationship for CV secondary stars \citep[e.g.][]{Knigge+2011}.

\item Analysis of the eclipse light curves. Similarly to the modelling of the spectral energy distribution (SED), this method provides both the effective temperature and the radius of the white dwarf, allowing to estimate the distance to the system \citep{Littlefair+2008,McAllister+2019}. 

\item $M_V - P_\mathrm{orb}$ relationship.
The radius of the accretion disc is a relatively fixed fraction of that of the white dwarf Roche lobe ($R_\mathrm{disc} \simeq 0.7 R_\mathrm{Roche-lobe}$, \citealt{Sulkanen+1981}), and scales with the mass ratio and the orbital separation, i.e. with $P_\mathrm{orb}$.  This, combined with the fact that disc outbursts occur when the accretion disc effective temperature rises to $\simeq 8000\,$K allows the use of dwarf nova outbursts as standard candles. This method was first introduced by \citet{Warner1987} and later refined by \citet{Patterson2011} using a sample of 46 CVs with good distance estimates. One of the main sources of uncertainty is the often unknown inclination of the system. 
\end{enumerate}
We identified 49 CVs with published pre-\gaia distance measurements of $d\le150$\,pc (Table \ref{tab:dist_pre-gaia}). For systems with multiple distance estimates, we considered the most reliable measurement following the order of the list above. 
\gaia re-identified all these 49 CVs, confirming that 28 of them are located within 150\,pc. The remaining 21 CVs are found to be located further than their pre-\gaia distances, the majority of which were estimated by modelling the red/infrared photometric contribution of the donor. This method systematically underestimates the distance to the systems (Figure~\ref{fig:pre-post-gaia}) owing to an unaccounted contamination by light from the accretion disc, which results in an overestimate of the brightness of the donor. In contrast, the distances estimated from the modelling of the SED in the ultraviolet are more accurate since the white dwarf is the dominant source of emission in this wavelength range. 

Since \gaia re-identified all the previously known CVs located within 150\,pc, we conclude that the completeness of our sample is not limited by the the ability of \gaia in obtaining astrometry of CVs but rather by the efficiency of the methods available to identify CVs in the first place. As discussed in Section~\ref{sec:sample}, the discovery of CVs has been strongly biased towards highly variable systems, such as dwarf novae with a high duty-cycle, and disfavoured the discovery of systems with both low mass transfer rates (WZ\,Sge dwarf novae, which have outburst recurrence times of the order of decades), and high mass transfer rates (novalike CVs in which the disc remains in a hot state). 

Discovering $>300$ new CVs, SDSS has demonstrated that the spectroscopic identification of CVs is largely independent of CV sub-type. In fact, follow-up of the SDSS CVs led to the unambiguous confirmation of a pile-up of intrinsically faint CVs near the period minimum \citep{Gaensicke+2009}. The SDSS CVs represent the currently least biased sample of CVs, and we make use of the $ugriz$ photometry to (i) estimate the completeness of our \gaia CV sample and (ii) search for CVs within 150\,pc that have so far been missed.

\subsection{Completeness of the 150\,pc CV sample}\label{sec:gaia_sdss}
Because the majority of known CVs fall within the colour space of quasars, we can make use of the highly efficient spectroscopic follow-up of SDSS quasar candidates to assess the completeness of the 150\,pc CV sample. The uniform spatial distribution of the 150\,pc CVs (Section~\ref{sec:150pc}) suggests that the properties of the local CV population are not evidently correlated with their location on the sky and that CVs are identified with equal success at low and high Galactic latitudes. Moreover, the typical CV scale height ($h = 280\,$pc) is much larger than the radius ($h = 150\,$pc) of the volume we are considering and consequently the correlation of the distribution of the 150\,pc CVs with the Galactic latitude is negligible.  
For this reason, although the SDSS footprint covers only one third of the sky and avoids the Galactic plane, the properties of the SDSS CVs can be safely extended to the whole sky and hence, to the 150\,pc \gaia CV sample.

\setlength{\tabcolsep}{0.2cm}
\begin{table*}
\caption{Summary of the cross-match between \gaia and SDSS and the relative sub-samples used to estimate the completeness in Section~\ref{subsec:completeness}.}\label{tab:xmatch}
\begin{tabular}{lr}
 \toprule
Description & Number of objects \\            
\midrule 
Total number of source in \gaia DR2 with $\varpi + 3\sigma_\varpi >= 6.66\,\mathrm{mas} $ & 2\,100\,094 \\
Number of source with reliable astrometry (Equation~\ref{eq:clean}) & 910\,187 \\
\quad of which in SDSS & 303\,723 \\
\quad \quad \quad of which in SDSS DR14 & 299\,501 \\
\quad \quad \quad of which in SDSS DR7 & 4222\\
\quad \quad of which within the Legacy and BOSS footprint & 155\,099 \\
\quad \quad \quad of which in the quasar colour space & 42\,823 \\
\quad \quad \quad \quad of which with similar colour to CVs & 5300 \\
\quad \quad \quad \quad \quad of which with a spectrum & 2811\\
\bottomrule 
\end{tabular}
\end{table*}

\subsubsection{Definition of the spatial and colour footprint followed-up by the SDSS quasar search}\label{subsec:colour_footprints}
With the aim to study extragalactic objects, SDSS acquired photometric observations for 7\,500\,square degree of sky and subsequently performed spectroscopic follow-up for a subset of the photometric objects. SDSS selected the spectroscopic quasar targets according to their colours \citep{Richards+2002}. Within the vicinity of a quasar in the four dimensional ($u-g$;$g-r$;$r-i$;$i-z$) colour space, all photometric objects have the same probability to be spectroscopically observed by the Legacy \citep{York+2000} and BOSS (Baryon Oscillation Spectroscopic Survey, \citealt{Dawson+2013}.

Consequently, the probability for an object to be observed spectroscopically depends on its location on the sky (i.e. whether it is located within the Legacy/BOSS footprint) and on its colour similarity to quasars. To define this parameter space we used the CasJobs\footnote{\url{https://skyserver.sdss.org/casjobs/}} service to query the PlateX table imposing \texttt{programName == "boss" || programName == "legacy"}. This returned the coordinates of the centres of the 4235 spectroscopic plates observed during the Legacy and BOSS surveys. Each spectroscopic plate has a field of view of $1.49$\degree\, in radius, therefore objects located within $\le 1.49$\degree\, from the plate centre that have (i) fiber magnitudes, i.e the flux contained within the aperture of a spectroscopic fiber, fainter than $15\,$mag in $g$ or $r$, and $14.5\,$mag in $i$ and (ii) $i < 19.1\,$mag, are all potential spectroscopic targets. These magnitude limits are imposed by the fixed 15~min exposure time of the SDSS observing strategy: objects brighter than the first set of limits saturate the SDSS detectors and objects fainter than $i=19.1$ are so numerous that they were not systematically followed-up spectroscopically.
We used the quasar catalogue from the SDSS DR7 \citep{Schneider+2010} to calculate the colour distance, i.e. the nearest neighbour distance, between each pair of quasar. We found that $\approx 99$\,per cent of them have another quasar within $\simeq 0.29\,$mag. We can hence assume that all objects found within $0.29\,$mag from a quasar are located within the colour space in which SDSS selected its spectroscopic targets and had therefore a chance to be observed spectroscopically.

\begin{figure}
\includegraphics[width=0.48\textwidth]{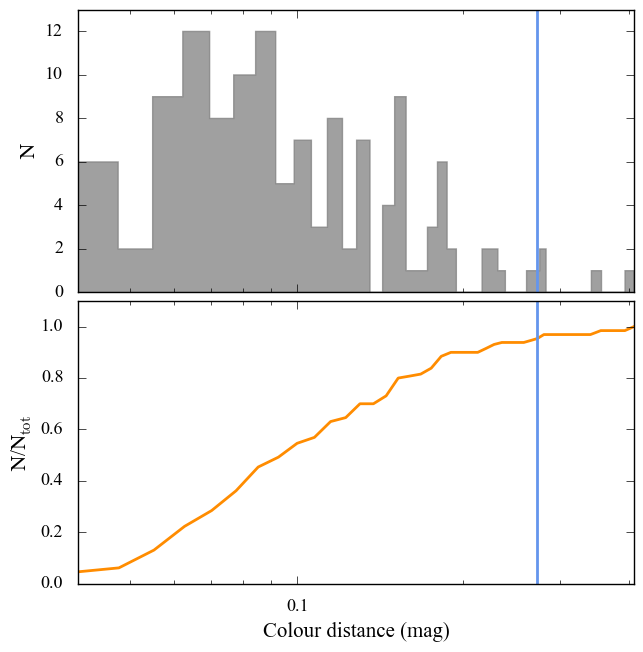}
\caption{Frequency distribution (top) and  cumulative distribution (bottom) of the colour distance in the four dimensional space ($u-g;g-r;r-i;i-z$) for the 130 SDSS CVs, showing that 95\,per cent of them have another CV within $\simeq 0.27\,$mag (vertical blue line).}\label{fig:color_dist}
\end{figure}

CVs occupy a sub-region of the colour space in in which quasars are found. This sub-region can be defined in an analogous fashion as done above for the case of quasars. We calculated the colours ($u-g; g-r; r-i; i-z$) of all known CVs with SDSS photometry. To avoid contamination from objects with poor photometry or objects mistakenly classified as CVs, we only considered SDSS sources that could be matched to CVs that were confirmed either by spectroscopy, or by a published orbital period measurement, and that have accurate (photometric errors $< 5\,$per cent) and clean SDSS photometry, i.e.:
\begin{equation}\label{eq:clean_sdss_phot}
\mathtt{clean\,=\,1~\&~mode\,=\,1~\&~type\,=\,6}.
\end{equation}
These flags ensure selection of stars (\texttt{type=6}) rather than galaxies and avoids blended photometry or multiple detections of the same source. We also required that the selected objects have reliable \gaia parallax and colours (Equation~\ref{eq:clean}).
The final sample contains high-quality photometry of 418 CVs, the majority of which are located further than 150~pc. In order to build a reference CV sample that is as representative as possible of the overall SDSS CV population, we computed the apparent magnitudes that SDSS CVs with $d > 150$~pc would have if they were located at $d = 150\,$pc. This allows us to remove those systems that, if they were closer, would have not been observed by SDSS owing to the bright limit in the target selection. Moreover, we only considered those CVs that fulfil the conditions listed above (location on the sky and colour similarity with quasars), which reduced the reference sample to 130 CVs (black crosses in Figure~\ref{fig:hr_all}).

We then defined a four-dimensional ``sphere'' in colour space, centred on each of these 130 CVs, and calculated the colour distance between each pair. Figure~\ref{fig:color_dist} shows that the distribution peaks at a colour distance of $\simeq 0.06\,$mag, however, to enclose 95\,per cent of the sample, we defined a colour radius of $\simeq 0.27\,$mag. 
All objects located within $\leq 0.27\,$mag from any of this 130 reference CVs are hence potential CV candidates. 
From their spectral completeness (i.e. the ratio between the objects with a spectrum and the total number of objects observed photometrically) combined with the fraction of spectroscopically observed objects drawn from that colour space that actually are CVs, one can then estimate how many CVs are expected to be found among the candidates. Comparing the expected number of CVs with that of known CVs within the reference spatial and colour footprint then provides the completeness of the sample.
In the following Section, we apply this method to the SDSS CVs in order to derive the completeness of the 150~pc \gaia CV sample.

\begin{figure}
\includegraphics[width=0.48\textwidth]{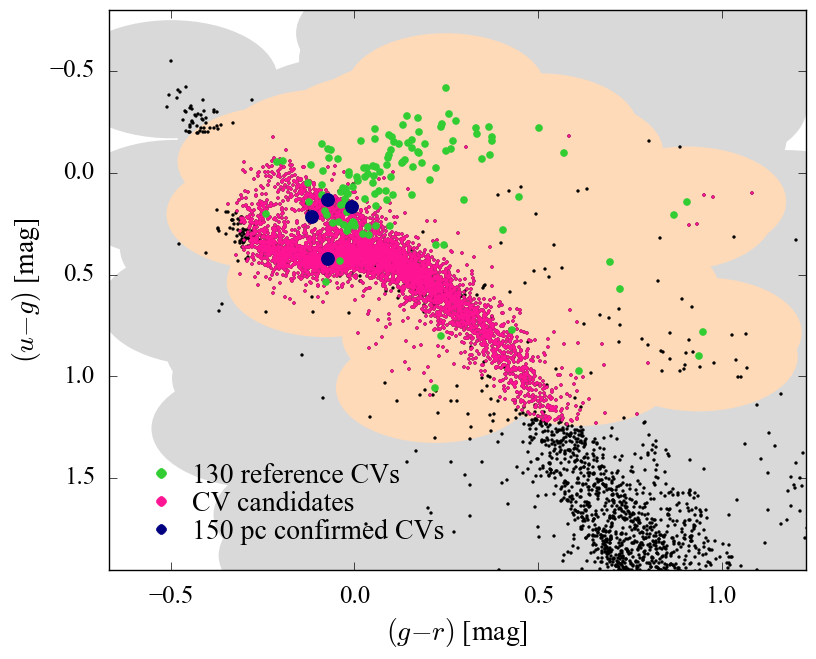}
\caption{Two-dimensional projection of the four-dimensional ($u-g$;$g-r$;$r-i$;$i-z$) colour space onto ($u-g$;$g-r$) of the sources within 150\,pc (black) that are found within the quasar colour space (grey). The four-dimensional colour space spanned by CVs has been defined by 130 reference CVs (green dots, Section~\ref{subsec:colour_footprints}), and the projection of that colours space onto (u-g,g-r) is shown by the light orange cloud. CV candidates contained within this four-dimensional colour space are shown in pink.
Among them, four confirmed CVs (dark blue) are known. Given the spectral completeness of $\simeq 53\,$per cent, $\approx$\,two more CVs are expected to be identified, implying an overall completeness of $\approx 71\,$per cent for the 150\,pc CV sample.}\label{fig:var_index}
\end{figure}

\begin{figure*}
\includegraphics[width=\textwidth]{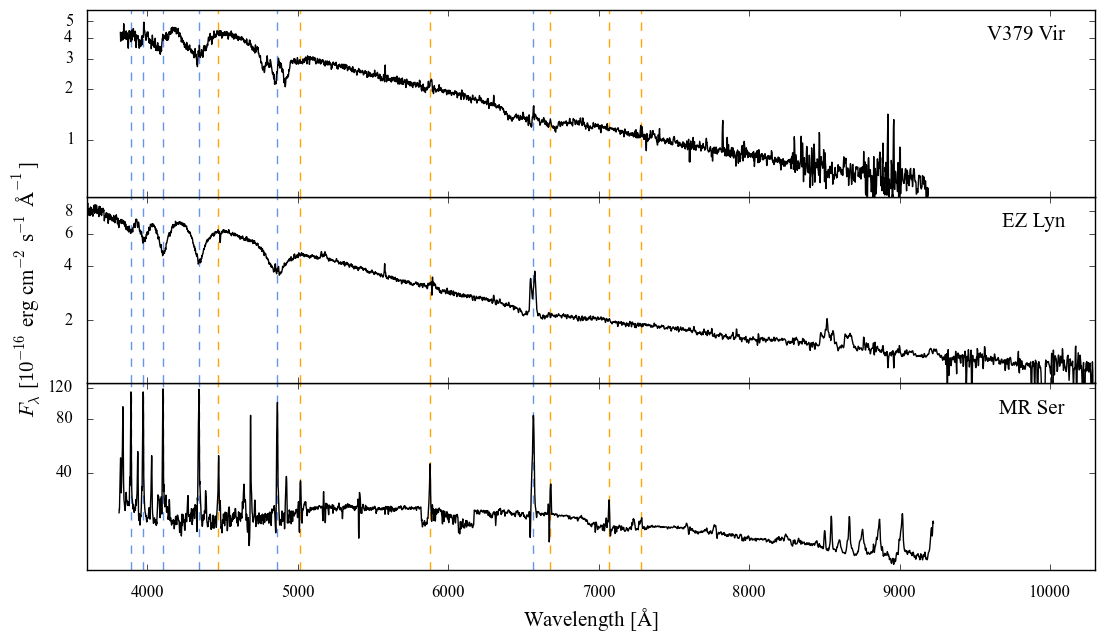}
\caption{SDSS spectra of the CVs in the 150~pc SDSS sample. The blue and the orange dashed lines highlight the position of the Balmer series and of some of the most common \ion{He}{i} lines, respectively.
EZ\,Lyn shows the typical spectral appearance of the low-accreting WZ\,Sge type CVs, with the white dwarf signature clearly recognisable in the broad Balmer line absorption features. Magnetic CVs undergo phases of high mass transfer (high states) alternated with phases in which accretion almost ceases (low states). Among the magnetic systems in the 150~pc sample observed by SDSS, V379\,Vir have been caught in a low state, while MR\,Ser has been observed in a high state.}\label{fig:sdss_spectra}
\end{figure*}

\begin{figure}
\includegraphics[width=0.5\textwidth]{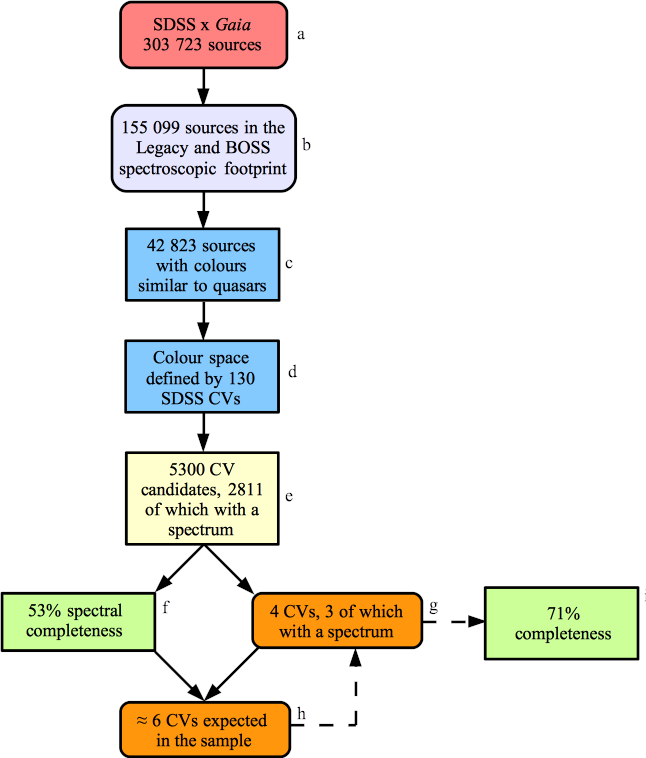}
\caption{Flowchart of the process used to estimate the completeness of the 150~pc sample. The starting point is the cross-match between \gaia and SDSS (a), which returns 303\,723 sources. A subsample of these, 155\,099 (b), falls within the footprint that the SDSS Legacy and BOSS surveys observed spectroscopically. Among them, 42\,823 have colours similar to quasars (c) and hence have all the same probability to have a spectrum in SDSS. 
Thanks to their colour similarity with the 130 SDSS CVs (d) that occupy a sub-region of the quasar colour space (see Section~\ref{subsec:colour_footprints}), 5300 CV candidates are identified, 2811 of which have a spectrum (e). This implies a spectral completeness of 53\,per cent (f). This sample contains four known CVs, three of which have a spectrum (g). We can hence estimate that $\approx$\,six CVs should be identify in total (h). Four CVs have already been found and we can conclude that the CV sample is $\simeq 71\,$per cent complete (i).}\label{fig:flow_chart}
\end{figure}

\subsubsection{Application to the 150~pc sample}\label{subsec:application}
We performed a cross-match between \gaia and SDSS\footnote{While the \gaia consortium already provides the cross-match with SDSS (gaiadr2.sdssdr9\_best\_neighbour), we found that $\simeq82\,000$ objects common to SDSS and \gaia\ are missing from that table, as well as $\simeq1000$ associations of spurious \gaia\ data with SDSS sources.}, querying the \gaia archive for all objects that, including their parallax uncertainties, are located within 150\,pc (Equation~\ref{eq:parallax_cut}), which returned $\simeq 2\,100\,000$ objects. Among them, many systems have inaccurate astrometry and $G_\mathrm{BP}-G_\mathrm{RP}$ colours due to faintness, blended double stars or other astrometric effects. Following the prescription from \citeauthor{Lindegren+2018} (\citeyear{Lindegren+2018}, their appendix C), we selected only those sources with the most reliable astrometry by applying the following cuts, 
\begin{equation}\label{eq:clean}
\begin{array}{lll}
\mathtt{astrometric\_excess\_noise} & < & 1\\
\mathtt{phot\_bp\_rp\_excess\_factor} & > & 1 + 0.015 \times \mathtt{bp\_rp}^2 \\
\mathtt{phot\_bp\_rp\_excess\_factor} & < & 1.3 + 0.06 \times \mathtt{bp\_rp}^2 \\
\end{array}
\end{equation}
which leaves $\simeq 910\,000$ objects with reliable \gaia colours and astrometry (grey dots in Figure~\ref{fig:hr_all}). 

We then queried the SDSS~DR14 archive using the CasJobs website and retrieved the SDSS coordinates, photometry, and Modified Julian Date of the observations for the closest objects within 30\arcsec\, to each \gaia entry. Using the \gaia proper motions, we calculated the coordinates of the \gaia sources at the epoch of the SDSS observations, and then performed a 2\arcsec\, radius cross-match with the SDSS objects, thus obtaining the best epoch-matched association between each \gaia entry and the SDSS~DR14 sample for a total of $\simeq 300\,000$ objects. 

Data releases later than DR7 provide more accurate photometry based on an improved background subtraction. As a side effect of this re-reduction of the photometry, some sources nearby bright stars that were present in DR7 are no longer included in the later data releases. To recover these lost sources, we applied the procedure outlined above also to the SDSS~DR7 catalogue, retrieving photometry for an additional $\simeq 4000$ sources that are no longer included in the subsequent data releases.

The final cross-match between the \gaia source within 150~pc and SDSS contains 303\,723 objects (Table~\ref{tab:xmatch}). 

Among these 303\,723 objects, 5300 (2811 of which have a spectrum) fall within the spatial and colour footprint of our SDSS CV reference sample defined above and are hence CV candidates (Figure~\ref{fig:var_index}). Four of these are known CVs (V379\,Vir, EZ\,Lyn, MR\,Ser, ASASSN-14dx), three of which (V379\,Vir, EZ\,Lyn and MR\,Ser) have a spectrum (Figure~\ref{fig:sdss_spectra}).

\begin{figure}
\includegraphics[width=0.48\textwidth]{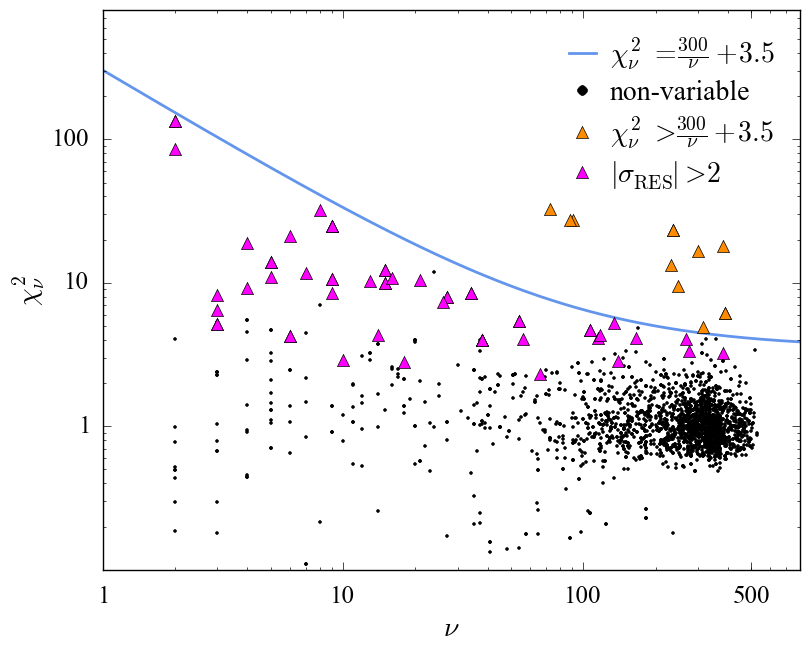}
\caption{Sample of 1807 hydrogen-atmosphere white dwarfs used to define the criteria for identifying variable systems. Systems above (orange points) or below the threshold (light blue) but with  $|\sigma_\mathrm{RES}| > 2$ (pink dots) are identified as variable.}\label{fig:DA_var}
\end{figure}

The spectra of these 2811 objects were then visually inspected, but no additional CV was identified. This is not surprising given the extensive search for CVs in the SDSS data that have been carried out in the past by P. Szkody and collaborators \citep{Szkody+2002, Szkody+2003, Szkody+2004, Szkody+2005, Szkody+2006, Szkody+2007, Szkody+2009, Szkody+2011}.
The spectral completeness of the CV candidate sample results $2811/5300 \simeq 53\,$per cent. Considering that there are three CVs with a spectrum, we can hence estimate that a total of $\approx$\,six CVs were expected to be identified. Since only four CVs have been detected, we can conclude that the sample is $\approx 71\,$per cent complete (Figure~\ref{fig:flow_chart}). 
This estimate of the completeness is subject to small number statistics, and we derive a more robust value making use of the much larger sample of CVs with CRTS light curves.

\subsubsection{Additional constraints from CRTS light curves}\label{subsec:variability}
\begin{figure*}
\includegraphics[width=0.85\textwidth]{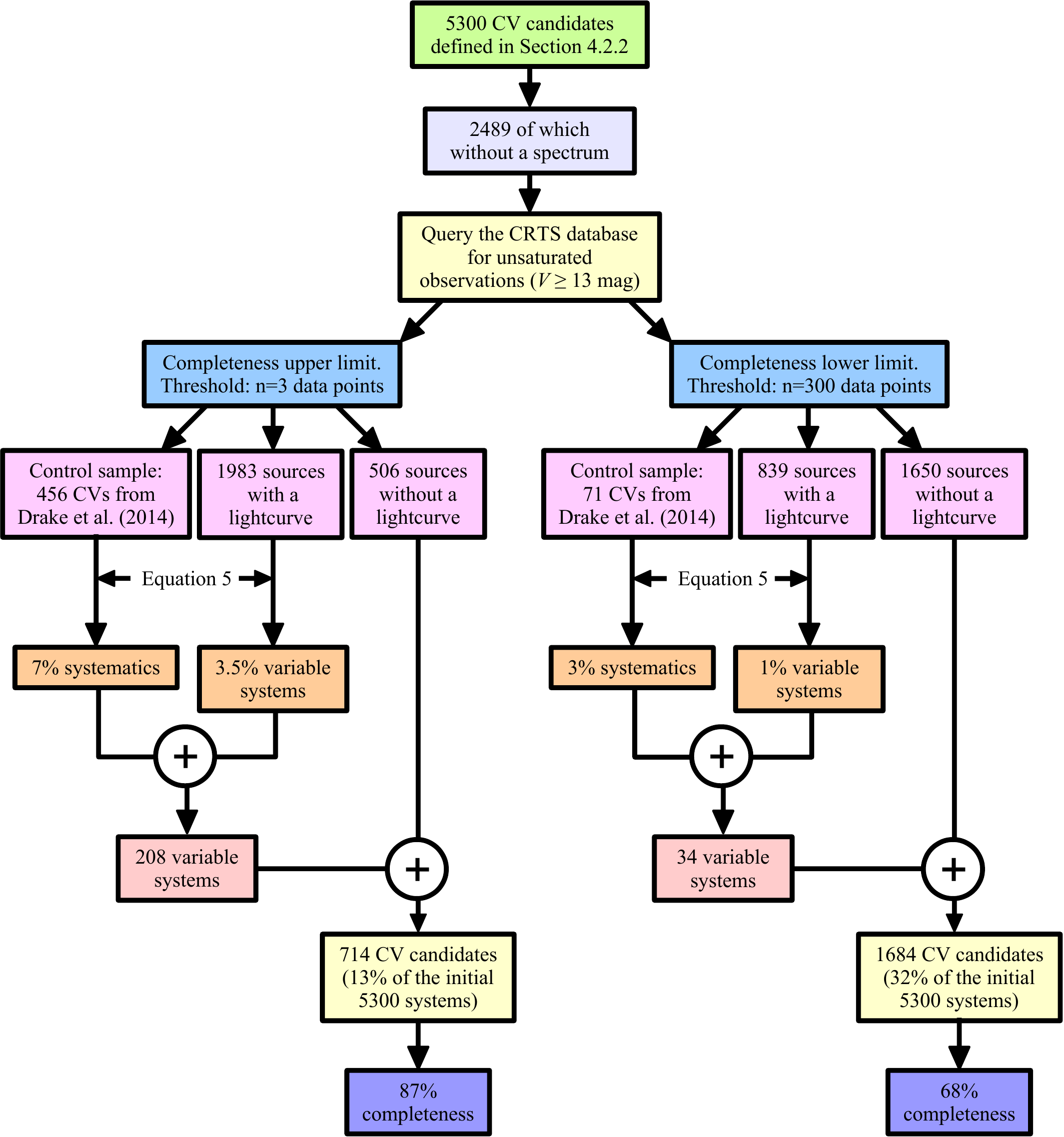}
\caption{Flowchart of the process used to estimate the completeness of the 150~pc sample using CRTS. The starting point is the 5300 CV candidates identified in Section~\ref{subsec:application}. 2489 of them do not have a spectrum and thus lack of additional information required to establishing their true nature. Using their CRTS light curves we establish the number of objects for which we are unable to unambiguously establish whether they are CVs, thus obtaining an upper and lower limit for the completeness of the 150\,pc CV sample.}\label{fig:flow_chart_crts}
\end{figure*}

Additional constraints on the completeness can be derived from the long-term light curves of the 2489 CV candidates identified in Section~\ref{subsec:colour_footprints} for which no SDSS spectroscopy is available, which provide the possibility to discover new CVs via the detection of their intrinsic variability. CRTS light curves are the most suitable for this task since, spanning more than 10 years, they are sensitive to the low mass transfer rate systems with disc outburst recurrence times of years \citep[e.g.][]{Breedt+2014}. Moreover, covering $\simeq 30\,000\,\mathrm{deg}^2$ of the sky between $-75\degree < \delta < 65\degree$ and $|b| \gtrsim 15\degree$ \citep{Drake+2009}, the CRTS footprint largely overlaps with that of SDSS, allowing us to combine the spectral completeness constraints derived in the previous section with the presence of possible variability.

Whereas the CRTS observing strategy aimed at obtaining data with $\simeq 2\,$week cadence, seasonal visibility and adverse weather condition can result in quite a different number of observations per field and therefore the CRTS data for a given object span a wide range in quality. Therefore it is necessary to define a criterion that identifies intrinsic variability opposed to possible scatter due to poor weather conditions, cosmic rays contamination and possible calibration issues. To this purpose, we used a sample of 1807 hydrogen-atmosphere white dwarfs from \citeauthor{Nicola+2019} (\citeyear{Nicola+2019}, flagged as ``DA'' in their table~4) which are located within 150\,pc, have been observed by CRTS, but are not variable (a small number of white dwarfs exhibit variability due to pulsations or rotations, though with amplitudes that are much lower than the variability seen among CVs).
Since these systems are not intrinsic variable, in an ideal data set, all data points of a light curve should be consistent with the median value within their uncertainties. Consequently the related reduced chi-squared $\chi^2_{\nu}$ should converge to $\approx 1$ as the number of degrees of freedom, $\nu = N-1$ (where $N$ is the number of observations), increases, whereas the standard deviation of the normalised residuals should be $|\sigma_\mathrm{RES}| < 2$.
We queried the CRTS database\footnote{\url{http://nesssi.cacr.caltech.edu/DataRelease/}} for the light curves of the white dwarf test sample. We considered only observations with $V \geqslant 13\,$mag to avoid saturated exposures and computed the $\chi^2_{\nu}$ associated with each light curve. We found that $\chi^2_{\nu}$ shows a much larger scatter than expected (Figure~\ref{fig:DA_var}), reflecting the presence of low-quality photometric observations which mimic the presence of intrinsic variability in the systems. 
Using those white dwarfs that appear intrinsically variable owing to the low quality of their CRTS observations (Figure~\ref{fig:DA_var}, orange and pink points), we defined the following two cuts that remove systems with poor CRTS data:
\begin{equation}\label{eq:variability_cut}
\chi ^2_{\nu} > \frac{300}{\nu} + 3.5\quad\mathtt{or}\quad|\sigma_\mathrm{RES}| > 2
\end{equation}
Three\,per cent of the objects satisfies the condition above and this fraction represents the intrinsic systematic uncertainty associate with this selection method.

To verify the quality of our criteria, we applied the same cuts to a control sample of 456 CVs from \cite{Drake+2014} for which CRTS light curves having at least three data points\footnote{Ideally, three data points represent a lower limit for the number of observations required to identify a CV candidate since two observations during quiescence allow to define a quiescent reference level with respect of which any significantly bright (even single) outlier can be indicative of the occurrence of a disc outburst.} with $V \geqslant 13\,$mag are available. Our method outlined above is able to identify as variable 93\,per cent of these CVs. The 7\,per cent of systems that we are unable to recover are those showing low variability or having a low number of observations ($N \lesssim 10$).

We then repeated the same exercise for the 2489 CV candidates without a spectrum (Table~\ref{tab:crts} and Figure~\ref{fig:flow_chart_crts}, left branch), since these are the objects for which the lack of additional information prevents us from establishing their true nature. As before, we queried the CRTS database for unsaturated exposures and we considered only light curves with three or more data points, which were available for 1983 objects.

Applying Equation~\ref{eq:variability_cut}, we find that 3.5\,per cent of them are variable. This fraction needs to be corrected for the fact that our method is unable to recover seven\,per cent of the 456 CVs from the control sample from \cite{Drake+2014}, thus resulting in a total fraction of 10.5\,per cent (i.e. 208) variable objects, which are possible CV candidates. These need to be complemented with the remaining 2489-1983=506 sources for which no information (neither a SDSS spectrum nor a CRTS light curve) is available and that could also hide a not yet identified CV. The total 714 systems, representing 13\,per cent of the 5300 candidates identified in the previous section, are the sources for which we are unable to unambiguously establish whether they are CVs and set an upper limit for the completeness of our sample to 87\,per cent.

A lower limit on the completeness can be derived imposing more stringent constraints on the CRTS light curve. Ideally, 200 points are in principle sufficient to identify short term variability (not necessarily related to disc instabilities, such as eclipses) in low mass accreting short period systems \citep{Parsons+2013}. However, in order to obtain a conservative lower limit on the completeness, we decided to impose a more stringent constraint, assuming that a minimum of 300 data points are necessary to identify variable systems.
In this case, our control sample is reduced to 71 CVs from \cite{Drake+2014} for which CRTS light curves with at least 300 unsaturated exposures are available and, with our selection criteria defined in Equation~\ref{eq:variability_cut}, we are able to recover as variable 97\,per cent of them.

\setlength{\tabcolsep}{0.2cm}
\begin{table*}
\caption{Summary of the different sub-samples used to estimate the completeness from CRTS light curves (Section~\ref{subsec:variability}).}\label{tab:crts}
\begin{tabular}{lr}
 \toprule
Description & Number of objects \\            
\midrule 
Source with colour similarity to CVs & 5300 \\
\quad of which without a spectrum & 2489 \\
\addlinespace[0.1cm]
\quad \quad of which with a CRTS light curve with more than three unsaturated data points & 1983 \\
\quad \quad \quad of which variable & 208 \\
\quad \quad of which without a CRTS light curve with more than three unsaturated data points & 506 \\
\addlinespace[0.1cm]
\quad \quad of which with a CRTS light curve with more than 300 unsaturated data points & 839 \\
\quad \quad \quad of which variable & 34 \\
\quad \quad of which without a CRTS light curve with more than 300 unsaturated data points & 1650 \\

\bottomrule 
\end{tabular}
\end{table*}

Among the 2489 CV candidates without a spectrum, 839 have a CRTS light curves matching the new requirement of 300 data points (Table~\ref{tab:crts} and Figure~\ref{fig:flow_chart_crts}, right branch) and one per cent of them results variable according to Equation~\ref{eq:variability_cut}. Accounting for the fact that, in this case, our method is unable to recover three\,per cent of CVs, results in a total fraction of four\,per cent variable objects, i.e. 34. Adding these to the 2489-839=1650 objects for which we have no information about (no SDSS spectra nor CRTS light curves) results in a total of 1684 CV candidates. These correspond to 32\,per cent of the initial 5300 candidates, implying a lower limit of 68\,per cent in the completeness of our sample.

Considering the upper and lower limits derived above, we conclude that our sample is ($77 \pm 10$) per cent complete, where the error is estimated as the sum of the maximum uncertainties in our method (seven per cent and three per cent respectively), representing the efficiency in recovering CVs as a function of the number of observations available.

\subsection{Summary on the completeness}\label{subsec:}
Using two independent methods, i.e. the count of objects with a spectrum (Section~\ref{subsec:application}) and the count of objects found to be variable (Section~\ref{subsec:variability}), we obtained two values for the completeness, $\approx 71\,$per cent and ($77\pm10$) per cent respectively, that are in good agreement among each other, thus supporting the validity of our result.

Because of the larger sample size of CVs used for the compute of the completeness carried out in Section~\ref{subsec:variability}, we adopt ($77 \pm 10$) per cent as the final value. As stated above, the CVs and CV candidates used in this analysis have only been selected according to their position on the sky and colours and hence the completeness estimate can be safely extended to the whole \gaia 150~pc CV sample. Given the ($77 \pm 10$) per cent completeness, we can estimate that $\approx 12$ CVs are still to be discovered within 150~pc.

The methods we employed to derive the completeness are based on the SDSS quasar target selection, and are biased against red, donor-dominated CVs. Moreover, such systems, if located within 150~pc are too bright to be selected as spectroscopic targets by SDSS owing to the saturation limit of the SDSS photometry. Nonetheless, it is reasonable to assume that all SU\,UMa and U\,Gem CVs within 150\,pc have already been identified since their brightening during their relatively frequent outbursts would have unlikely been missed by the many time-domain surveys and amateur observers that regularly scan the sky for transient events (see e.g. \citealt{Breedt+2014} for a case study of CRTS). An example is U\,Gem, the prototype for its class, which is known since 1855 \citep{Pogson1857} and has been been observed photometrically by SDSS, but not spectroscopically owing to its brightness ($g = 14.6\,$mag).

Instead, we cannot exclude that some WZ\,Sge systems, novalike CVs and polars within 150\,pc remain to be identified: ASASSN-14dx, a WZ\,Sge star located at $d=81.0\pm 0.3\,$pc with a quiescent magnitude of $V \simeq 16.2\,$mag \citep{Thorstensen+2016}, and TCP\,J21040470+4631129, another WZ\,Sge-type CV located at $d=109\pm2\,$pc with a quiescent magnitude of $V \simeq 17.7\,$mag (Atel\,\#12936), have only been discovered in 2014 and 2019, respectively, following a dwarf nova outburst.
We can conclude that the missing systems are most likely those showing low or no variability, such as (i) low mass transfer rate systems with outburst recurrence times of decades, (ii) high mass transfer rate systems (i.e. novalikes) with stable, hot accretion discs and (iii) strongly magnetic disc-less systems that do not experience disc outburst, which have been in a low-state during the \textit{ROSAT} All Sky X-ray survey, and other serendipitous X-ray observations.

As a final remark, it is important noticing that the colour footprint in which we searched for CV candidates (Figure~\ref{fig:var_index}) includes the region in which these low variability systems are expected to be located and therefore the methods we used to derive the completeness are optimally accounting for those objects that most likely are still missing in the 150\,pc CV sample.

\begin{figure*}
\includegraphics[width=0.7\textwidth]{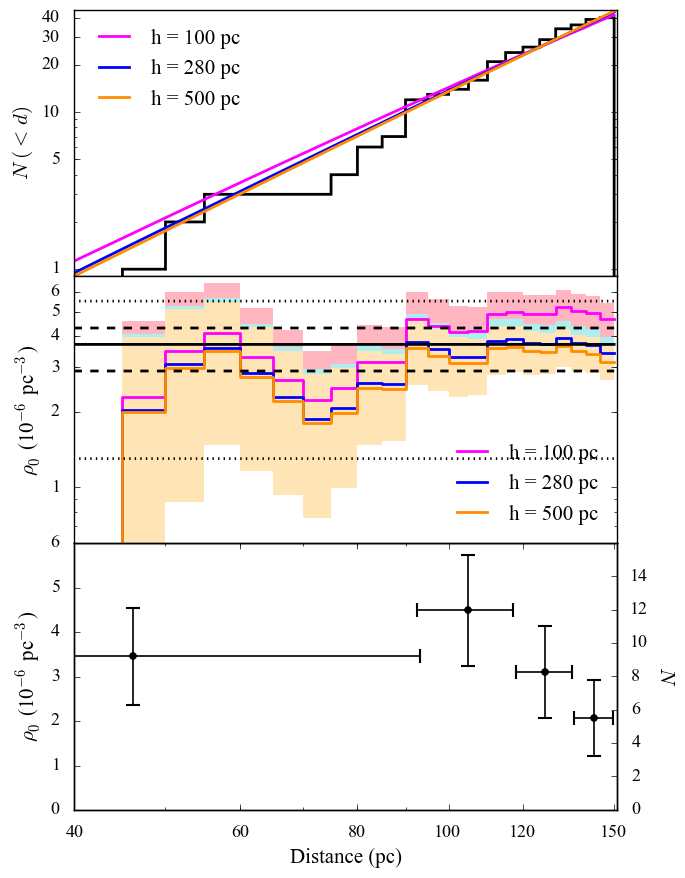}
\caption{\emph{Top:} cumulative distribution of the 150\,pc CV sample as a function of the distance (black line) in comparison with the prediction by the Galactic model assuming different scale heights (coloured lines as indicated). \emph{Middle:} $\rho_0$ as a function of distance for different scale heights. The black solid line corresponds to our conservative value of the CV space density $\rho_0 = (3.7^{+0.6}_{-0.8}) \times 10^{-6}\,\mathrm{pc}^{-3}$ for $h = 280\,$pc. The dashed and dotted lines represent the relative $1\,\sigma$ and $3\,\sigma$ uncertainties, respectively. \emph{Bottom:} CV space density in four bins of equal volumes. The error bars report the Poissonian uncertainties.}\label{fig:rho}
\end{figure*}

\section{Space density}\label{sec:space_density}
To determine the space density of CVs, we assumed a Galactic model following the prescription of \citet{Pretorius+2007a} and approximated the Galaxy as an axisymmetric disc, with no halo, no bulge, no spiral structure and no thick disc. We assumed the following density profile:
\begin{equation}
\rho = \rho_0 \exp \left( -\frac{|z|}{h} \right)
\end{equation}\label{eq:gal_model}
where $z$ is the distance above the Galactic plane, $\rho_0$ the space density in the mid-plane, and $h$ the scale height of the CV population. We ignored any possible dependencies from the radial distance from the Galactic centre since they are negligible within the volume we are considering.

$\rho_0$ is then calculated as the ratio between the number of CVs ($N_\mathrm{CV}$) found in the effective volume ($V_\mathrm{eff}$) for a given Galactic model (i.e. for a given scale height) and the effective volume itself. 
One of the unknowns in the determination of the space density is the scale height of the CV population. As discussed in Section~\ref{sec:150pc}, $h$ depends on the age of the population, and with the small number of CVs within 150\,pc, it is not possible to independently measure $h$.
We therefore assumed $h = 100, 280$ and $500\,$pc and determined $\rho_0$ using a Monte Carlo approach to evaluate the effective volume enclosed within a given distance (middle panel of Figure~\ref{fig:rho}). For our distance limit of 150~pc, we found a variation of $\simeq 30\,$per cent of $\rho_0$ for the two extreme cases of $h$ ($\simeq 4.8 \times 10^{-6}\,\mathrm{pc}^{-3}$ for $h = 100$\,pc, and $\simeq 3.4 \times 10^{-6}\,\mathrm{pc}^{-3}$ for $h = 500$\,pc respectively, see Table~\ref{tab:space_density}).
However, the 150\,pc CV sample is dominated by old CVs and hence the higher values we assumed for $h$ are more likely to be more representative of the properties of the observed sample.

\begin{figure}
\includegraphics[width=0.48\textwidth]{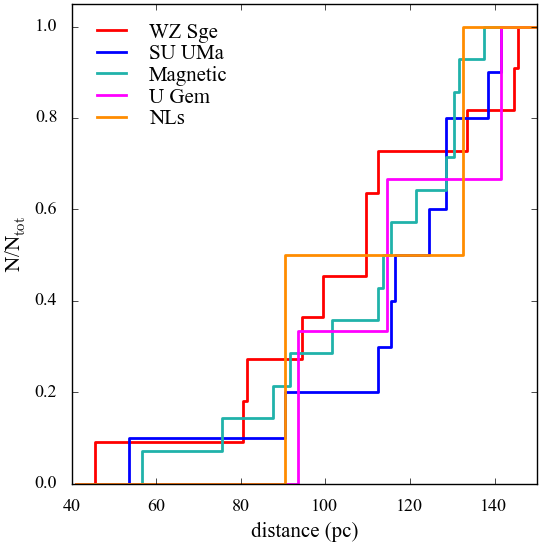}
\caption{Cumulative distribution of the 150~pc CVs as a function of the distance for the different subtypes. The different distributions shows all similar trends, suggesting the absence of clear selection effects in the 150~pc sample.}\label{fig:histo_type}
\end{figure}

In order to investigate for the presence of possible biases as a function of distance, we divided the 150\,pc volume in four bins of equal volume (bottom panel of Figure~\ref{fig:rho}). Within the uncertainties $\rho_0$ remains constant in the four bins, although its value decreases in the outer bin, reflecting the presence of possible detection biases. However, the cumulative distributions of the different CV sub-types (Figure~\ref{fig:histo_type}) show all the same trend suggesting that, whether present, these detection biases affect all CV sub-classes in the same way.

\setlength{\tabcolsep}{0.15cm}
\begin{table}
\centering
\caption{CV space densities for different scale heights. Both the values and the number of systems ($N_\mathrm{CV}$) derived from the analysis of the \gaia data and the corresponding ones corrected for the completeness of the sample are reported.}\label{tab:space_density}
\begin{tabular}{lcccc}
\toprule
\hspace{2.7cm}$d_\mathrm{lim}$ (pc) & & 
\multicolumn{2}{c}{150} \\ 
\cmidrule{3-4}
\diagbox[height=2.5\line]{$h$ \\ (pc)}{Completeness \\ $N_\mathrm{CV}$} & &
\multicolumn{1}{c}{\raisebox{0.35\normalbaselineskip}[0pt][0pt]{\makecell{77\%\\ 42}}} & 
\multicolumn{1}{c}{\raisebox{0.35\normalbaselineskip}[0pt][0pt]{\makecell{100\%\\ 54}}} \\
\midrule
& & \multicolumn{2}{c}{$\rho_0 ~[\times 10^{-6}$~\pc]} \\
\cmidrule{3-4}
  100 & & 
  $4.8^{+0.8}_{-1.0}$ & $6.3^{+0.9}_{-1.1}$  \\
  \addlinespace[0.2cm]
  280 & & 
  $3.7^{+0.6}_{-0.8}$ & $4.8^{+0.6}_{-0.8}$ \\
 \addlinespace[0.2cm] 
  500 & & 
  $3.4^{+0.5}_{-0.6}$ & $4.3^{+0.5}_{-0.7}$ \\
\bottomrule  
\end{tabular}
\end{table}

Considering the volume enclosed within 150\,pc, and accounting for the Poisson uncertainties, a conservative measurement of the space density of known CVs results $\rho_0 = (3.7^{+0.6}_{-0.8}) \times 10^{-6}\,\mathrm{pc}^{-3}$ for $h = 280\,$pc (solid line in the middle panel Figure~\ref{fig:rho}). The related $3\,\sigma$ uncertainties (dotted lines in the middle panel Figure~\ref{fig:rho}) also include the uncertainty due to the unknown scale height of the Galactic CV population. Accounting for the possibility that the \gaia CV sample is only 77\,per cent complete, would imply a space density of $\rho_0 = (4.8^{+0.6}_{-0.8}) \times 10^{-6}\,\mathrm{pc}^{-3}$ for $h = 280\,$pc, which would still be consistent with the conservative value derived above without introducing any correction for incompleteness.

Prior to our \gaia-based analysis, the most reliable space density had been estimated using an X-ray selected sample of CVs, $\rho_0 =4{+6\atop-2}\times 10^{-6}\,\pc$ \citep{Pretorius+2012}. 
\citet{Schwope2018} recently re-visited this sample, making use of the \gaia distances, and derived $\rho_0 < 5.1\times 10^{-6}\pc$.
Both studies assumed that this magnitude-limited sample of X-ray selected CVs is complete and representative of the intrinsic population, leaving systematic uncertainties of a factor two in the space density measurement \citep{Pretorius+2012}.

Our measurement, $\rho_0 = (4.8^{+0.6}_{-0.8}) \times 10^{-6}\,\mathrm{pc}^{-3}$, is in good agreement with these previous estimates. However, using a volume-limited sample and the accurate astrometry of \gaia, we were able to reduce the uncertainty on the CV space density by an order of magnitude. 

\section{Comparison with the models of CV evolution}\label{sec:test}
\subsection{The intrinsic population}\label{subsec:intrinsic}
The volume-limited sample of CVs obtained from \gaia provides, for the first time, a direct insight into the intrinsic properties of the Galactic population of CVs and allows a direct comparison between the observations and theoretical predictions. 

\begin{figure}
\includegraphics[width=0.48\textwidth]{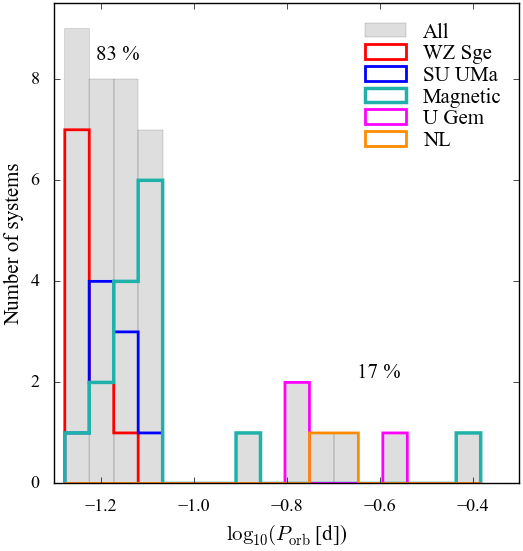}
\caption{Period distribution for the CVs in the 150~pc sample. The majority of the systems are found below the period gap and, accounting also for the two WZ\,Sge-type stars without an orbital period determination, make up for $\simeq 83\,$per cent of the observed systems.}\label{fig:histo_period}
\end{figure}

Most models of CV evolution predict that $\simeq 99\,$per cent of the present day CVs should be found below the period gap, with a large fraction of them ($\simeq 40-70\,$per cent) having already evolved through the period minimum \citep{Kolb1993,Howell+2001,Goliasch-Nelson2015}.
The results from \gaia show instead a different picture, with $(83 \pm 6)\,$per cent of the CVs in the 150\,pc sample below the period gap and $(17\pm 6)\,$per cent above (Figure~\ref{fig:histo_period}). More importantly, the \gaia 150\,pc CV sample contains only three plausible candidate period bouncers, GD\,552 \citep{Unda-Sanzana2008}, SDSS\,J102905.21+485515.2 \citep{Thorstensen+2016} and 1RXS\,J105010.3--140431 \citep{Patterson2011,Pala+2017}, with WZ\,Sge, V455\,And and EZ\,Lyn being three additional weaker candidates (they all have brown-dwarf companions, but are right at the period minimum). Thus seven to at most 14 per cent of the 150\,pc CVs below the gap are period bouncers, a much smaller fraction than predicted by the population models.
This discrepancy could reflect the selection biases discussed above, i.e. it could be possible that a number of period bouncers have not yet been identified because of their low quiescent variability and long outburst recurrence time. 
Assuming that the \gaia 150\,pc CV sample is 77\,per cent complete (Section~\ref{subsec:completeness}), $\simeq 12$ CVs are still to be identified within 150\,pc. The most favourable case in which all these are period bouncers would results into a fraction of $(31\pm7)$\,per cent, bringing the observations into a marginal agreement with the prediction by \citet{Goliasch-Nelson2015}, $\simeq 40\,$per cent, although the fraction of CVs above, ($13 \pm 5)$\,per cent, and below the gap, ($87 \pm 5$)\,per cent, would still be quite different from the theoretical predictions.

\setlength{\tabcolsep}{0.22cm}
\begin{table*}
\caption{Literature effective temperatures and masses for the CV white dwarfs in the 150\,pc sample, ordered according to their orbital periods.}\label{tab:cvwd_teff}
\begin{tabular}{@{}lcccccc@{}}
 \toprule
 System         & $P_\mathrm{orb}$ (min) & $T_\mathrm{eff}$ (K) & $\langle \dot{M} \rangle$ ($10^{-11} \mathrm{M}_\odot/\mathrm{yr}^{-1}$)& $M_\mathrm{WD}$ (M$_{\odot}$) & Comment & Reference\\
\midrule
GW\,Lib  & 76.78 & $ 16\,995 \pm 812 \downarrow$ & $13 \pm 3$ & $0.84 \pm 0.02$ & Pulsating, brightenings & 17, 21 \\  
BW\,Scl      & 78.23 & $15\,480 \pm 900$ & $10 \pm 7$ & -- &  & 16\\ 
V627\,Peg & 78.51 & $16\,292 \pm 753$ & $13 \pm 9$ & --  & & 17 \\  
V455\,And & 81.08   & $ 11\,799 \pm 750 $  & $4 \pm 2$ & -- & Eclipsing, pulsating & 8\\ 
WZ\,Sge      & 81.63    & $14\,900 \pm 250$ & $7.4 \pm 1.3$ & $0.85 \pm 0.04$ & & 1, 2\\ 
V355\,UMa    & 82.52    & $12\,924 \pm 1250$ & $5 \pm 4$ & -- & Pulsating & 33 \\
EZ\,Lyn & 84.97 & $13\,595 \pm 99$ & $6 \pm 4$  & -- & Pulsating, brightenings & 27\\ 
1RXS\,J105010.3--140431 & 88.56 & $11\,622 \pm 277$ & $3 \pm 2$  & -- & Period bouncer candidate & 17\\  
V2051\,Oph & 89.90 & --  & -- & $0.78 \pm 0.06$ & Eclipsing & 19, 28\\ 
VY\,Aqr & 90.85 & $ 15\,148 \pm 100 $ & $10 \pm 6$ & -- & & 23\\ 
OY\,Car       & 90.89 & $ 15\,000 \pm 2000$ & $8 \pm 5$ & $0.84 \pm 0.04$ & Eclipsing & 11, 12\\ 
EX\,Hya        & 98.26   &  -- & -- & $0.790 \pm 0.026$ & Magnetic & 5,6,7,32 \\ 
VV\,Pup & 100.44 & $ 12\,251 \pm 600 $ & $4 \pm 3$ & -- & Magnetic & 20\\ 
V834\,Cen & 101.52 &  $14\,927 \pm 900$ & $10 \pm 6$ & -- & Magnetic & 20\\ 
GD\,552      & 102.73 & $ 11\,118 \pm 400 $ & $2.9 \pm 1.9$ & -- & Period bouncer candidate & 9\\ 
HT\,Cas & 106.05 & $ 14\,000 \pm 1000$ & $22 \pm 8$ &   $0.61 \pm 0.04$ & Eclipsing & 25, 26\\ 
VW\,Hyi       & 106.95 & $ 20\,000 \pm 1000$ & $50 \pm 34$&  $0.71^{+0.18}_{-0.26}$ & & 3, 4\\ 
Z\,Cha         & 107.28 & $15\,700 \pm 550$ & $10 \pm 3$&  $ 0.84\pm 0.09$ & Eclipsing & 29, 30\\
MR\,Ser & 113.47 & $14\,816 \pm 900 $ & $9 \pm 6$&  -- & Magnetic & 20\\ 
BL\,Hyi & 113.64 & $ 13\,818 \pm 900 $ & $7 \pm 5$&  -- & Magnetic  & 20\\
ST\,LMi  & 113.89 & $ 11\,005 \pm 500 $ & $2.8 \pm 1.8$& -- & Magnetic & 20\\
AM\,Her       & 185.65 & $ 19\,800 \pm 700 $ & $33 \pm 17$ & $0.78^{+0.12}_{-0.17}$ & Magnetic & 10\\ 
IP\,Peg & 227.82 & -- & -- &$1.16 \pm 0.02$ & Eclipsing & 24\\
U\,Gem      & 254.74 & $ 30\,000 \pm 1000$ & $31 \pm 6$ & $1.2 \pm 0.05$ & & 14, 15\\  
IXVel       & 279.25  & -- & -- & $0.8 \pm 0.2$  & & 31\\
SS\,Cyg & 396.19 & -- &  -- & $0.81 \pm 0.19$ & & 22\\
AE\,Aqr      & 592.78 & -- &  -- & $ 0.63 \pm 0.05 $ & Magnetic, evolved donor & 13\\
\bottomrule
\end{tabular}
\begin{tablenotes}
\item \textbf{References.} (1) \cite{Sion+1995}, (2) \cite{Steeghs+2007}, (3) \cite{Gaensicke+1996}, (4) \cite{Smith+2006}, (5) \cite{Eisenbart+2002}, (6) \cite{Belle+2003}, (7) \cite{Beuermann+2008}, (8) \cite{Araujo-Betancor+2005}, (9) \cite{Unda-Sanzana2008}, (10) \cite{Gaensicke+2006}, (11) \cite{Horne+1994}, (12) \cite{Littlefair+2008}, (13) \cite{Echevarria+2008}, (14) \cite{Long+2006}, (15) \cite{Echevarria+2007}, (16) \cite{Gaensicke+2005}, (17) \cite{Pala+2017}, 
(19) \cite{Baptista+1998}, (20) \cite{Araujo-Betancor+2005}, (21) \cite{vanSpaandonk+2010}, (22) \cite{Bitner+2007}, (23) \cite{Sion+2003}, (24) \cite{Copperwheat+2010}, (25) \cite{Feline+2005}, (26) \cite{Horne+1991}, (27) \cite{Szkody+2013}, (28) \cite{Saito+2006}, (29) \cite{Robinson+1995}, (30) \cite{Wade+1988}, (31) \cite{Neustroev+2011}, (32) \cite{Suleimanov+2019}, (33) \cite{Szkody+2010_v355uma}
\item \textbf{Notes.} For BW\,Scl, V455\,And, V355\,UMa, EZ\,Lyn, VY\,Aqr, VV\,Pup, V834\,Cen, GD\,552, MR\,Ser, BL\,Hyi and ST\,LMi, the white dwarf $T_\mathrm{eff}$ reported in the literature has been determined via spectroscopic analyses assuming $\log g = 8$. However, this assumption does not reflect the average mass of CV white dwarfs $M_\mathrm{WD} \simeq 0.8\,\mathrm{M}_\odot$, i.e. $\log g = 8.35$. For these systems, the $T_\mathrm{eff}$ reported in this table have been re-computed for $\log g = 8.35$ using equation~1 from \citet{Pala+2017}.
\end{tablenotes}
\end{table*}

This observed disagreement can be potentially resolved by accounting for the presence of additional AML mechanisms besides the mere MB and GWR. \citet{Schreiber+2016} proposed a model in which an empirical consequential AML (eCAML), i.e. a mechanism of AML arising from the mass transfer process itself, is generated by the friction between the secondary star and the white dwarf ejecta during nova explosions. The eCAML could lead to dynamically unstable mass transfer in CVs hosting low-mass white dwarfs, which would not survive as semi-detached binaries but would merge into single objects \citep[see also][]{Nelemans+2016}. As a consequence, the fraction of CVs above and below the period gap would become, respectively, $\simeq 85\,$per cent and $\simeq 15\,$per cent, and would be in good agreement with what derived from the analysis of the \gaia 150\,pc CV sample, $(83 \pm 6)\,$per cent and $(17 \pm 6)\,$per cent, even when accounting for incompleteness, $(87 \pm 5)\,$per cent and $(13 \pm 5)\,$per cent.
The space density predicted by the eCAML model, $\rho_0 \lesssim 2 \times10^{-5}\,\pc$ \citep{Belloni+2018}, is about a factor 4 higher than the space density derived from the study of the \gaia 150\,pc sample, $\rho_0 = (4.8^{+0.6}_{-0.8}) \times 10^{-6}\,\mathrm{pc}^{-3}$. 
This difference likely reflects the general uncertainties on the parameters (such as initial mass ratio distribution, initial separation distribution, initial binary fraction, common-envelope and magnetic braking efficiency) employed in binary population synthesis studies.
Moreover, the space density derived by \cite{Belloni+2018} has been derived from a CV population consisting of 80\,per cent period bouncers, while the fraction of such systems is only seven per cent in the observed 150\,pc sample. Considering only systems that have not evolved through the period minimum yet, the space density predicted by the eCAML model, $\rho_0 \approx 4.5^{+4.5}_{-2.3} \times10^{-6}\,\pc$ \citep{Diogo+2019}, perfectly agrees with the observation. 
This implies that either the large fraction of period bounce CVs still has to be identified (although this is not likely the case, see Section~\ref{subsec:application}), or that the current models fail to properly describe the post period minimum evolution of CVs. Alternatively, it is possible that the time scales required for a CV to evolve to the period minimum are much longer than current models (including eCAML) suggest.

\begin{figure*}
\includegraphics[width=\textwidth]{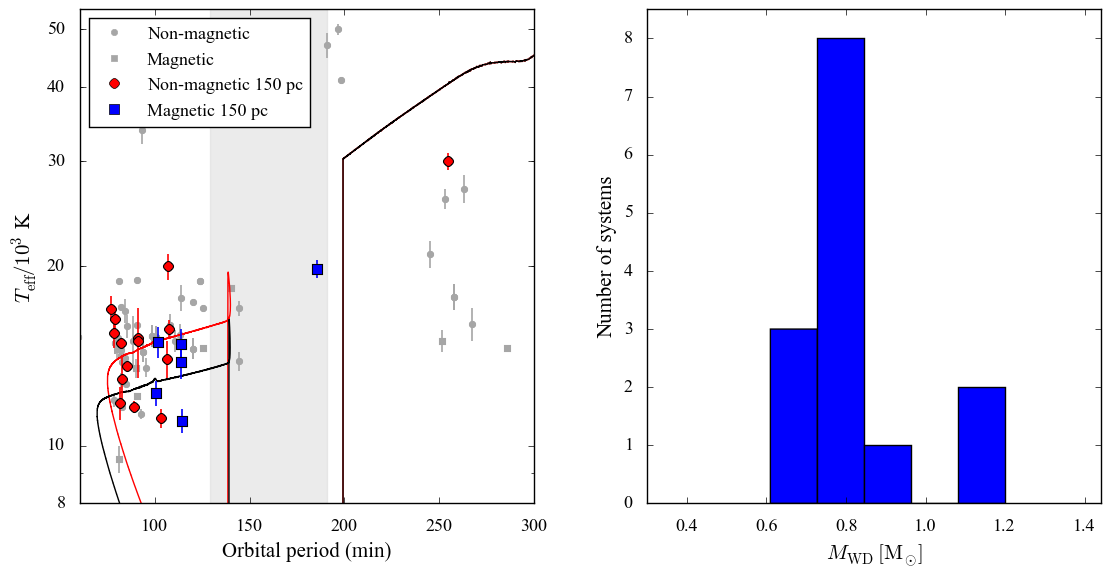}
\caption{\emph{Left:} effective temperature for magnetic (squares) and non-magnetic (circles) CV white dwarfs from \citet{Townsley+2009} and \citet{Pala+2017} (grey). Magnetic and non-magnetic systems within 150~pc from Table~\ref{tab:cvwd_teff} are shown in blue and red, respectively.
The period gap is highlighted by the grey band. The solid lines represent the evolutionary tracks from \citet{Pala+2017} for a typical CV ($M_\mathrm{WD} = 0.8\,\mathrm{M}_\odot$, with an initial secondary mass of $M_2 = 0.65\,\mathrm{M}_\odot$ and an initial $P_\mathrm{orb} = 12\,$h) in which AML is driven by both MB and GWR above the period gap while, below the period gap, only GWR (black) or GWR plus a residual MB (red) drive the evolution of the system. \emph{Right:} white dwarf mass distribution for the CVs in the 150~pc sample as reported from the literature (Table~\ref{tab:cvwd_teff}).}\label{fig:temp_mass_distr}
\end{figure*}

Finally, we note in passing that the 150\,pc sample contains some of the most peculiar CVs known:
\begin{itemize}
\item AR\,UMa is the polar with the highest magnetic field, $\langle B \rangle = 230\,$MG \citep{Schmidt+1996}. This suggests that strong magnetic fields are probably not as rare as thought \citep{Ferrario+2015} and underlines the urgency to better understand the origin of magnetic CVs.
\item AE\,Aqr is a post thermal time scale mass transfer system (see Section~\ref{subsec:evolved_donors}). It is also an IP, with the fastest spinning white dwarf, $P_\mathrm{spin} = 33\,$s \citep{Patterson+1980} among all CVs. This rapid spin results in a propeller mechanism which prevents the mass lost from the donor to reach the white dwarf surface, and is instead expelled in the surrounding space.
\item EZ\,Lyn is characterised by quasi-periodic brightening events superimposed to a sinusoidal photometric modulations \citep{Zharikov+2008,Zharikov+2013} that, in the case of its twin system SDSS\,J123813.73-033933.0, have been interpreted as the interplay between spiral arms and small amplitude thermal instabilities in the accretion disc \citep{Pala+2019}. Similar behaviour, although less periodic, has been observed also in GW\,Lib, where it could be associated with fluctuation in the mass accretion rate \citep{Odette}.
\item V445\,And, ``The CV that has it all'', displays a grazing eclipse, permanent superhumps, non-radial white dwarf pulsations and a spectroscopic period much longer ($P_\mathrm{spec} \simeq 3.5\,$h) than the orbital one ($P_\mathrm{orb} \simeq 81.08\,$min, \citealt{Araujo-Betancor+2005}), the origin of which is still not understood.
\end{itemize}
The prevalence of several of such peculiar systems suggests that the Galactic population of CVs is intrinsically very variegated and that these systems do not represent exceptional cases. The unexpected behaviour of systems like EZ\,Lyn and V455\,And suggests that the physics describing the accretion process is still far from being completely understood. Moreover, the existence of systems such as AR\,UMa and AE\,Aqr is not accounted for by the model of CV evolution and provides another clue of the incompleteness of the current theories describing the evolution of close interacting binaries. 

\subsection{Mass accretion rates and white dwarf masses}
Mass accretion results in compressional heating of the white dwarfs in CVs \citep{Sion1995,Townsley+04} and, therefore, the white dwarf effective temperature ($T_\mathrm{eff}$) provides a direct measurement of its secular mean accretion rate ($\langle\dot{M}\rangle$, \citealt{Townsley+03}).
The different efficiencies of MB and GWR in removing angular momentum from the binary orbit (Section~\ref{sec:intro}) cause long period CVs to have  $\langle\dot{M}\rangle$ about one order of magnitude higher compared to those of short period CVs. Consequently, long period systems should host hotter white dwarfs compared to short period systems and hence $T_\mathrm{eff}$ measurements provide a direct insight into the evolutionary stage of the systems \citep{Townsley+2009,Pala+2017}.

Among the 42  CVs found within 150\,pc, 21 have a published $T_\mathrm{eff}$ that can be considered reliable\footnote{Following the prescription by \citet{Townsley+2009}, a $T_\mathrm{eff}$ measurement is considered reliable when it has been derived (i) from the analysis of an ultraviolet spectrum in which the white dwarf signature has been unambiguously identified from the detection of a broad Ly$\alpha$ absorption profile and, possibly, sharp absorption metal lines, or (ii) from the analysis of the eclipse light curve in which both the white dwarf ingress and egress have been clearly detected.} and we can hence estimate their $\langle\dot{M}\rangle$ using equation~2 from \citet{Townsley+2009}. 
Table~\ref{tab:cvwd_teff} lists the corresponding $T_\mathrm{eff}$ values as compiled from \citet{Townsley+2009} and \citet{Pala+2017}, and includes four additional systems  EZ\,Lyn, GD\,552, Z\,Cha and V355\,UMa. For the systems for which the $T_\mathrm{eff}$ from the literature has been determined via spectroscopic analyses assuming $\log g = 8$, we re-computed the corresping $T_\mathrm{eff}$ for $\log g = 8.35$ using equation~1 from \citet{Pala+2017}, in order to reflect the average observed mass of CV white dwarfs $M_\mathrm{WD} \simeq 0.8\,\mathrm{M}_\odot$, i.e. $\log g = 8.35$. The  errors on $\langle\dot{M}\rangle$ have been derived accounting for the uncertainties on both $T_\mathrm{eff}$ and the white dwarf mass ($M_\mathrm{WD}$). For EZ\,Lyn, VY\,Aqr and V355\,UMa we adopt ten per cent of their $T_\mathrm{eff}$ values as uncertainty, as no errors on $T_\mathrm{eff}$ were published. For the systems without a mass measurement, we assumed $M_\mathrm{WD} = 0.83 \pm 0.23\,\mathrm{M}_\odot$, corresponding to the average mass of CV white dwarfs \citep{Zorotovic}.

The left panel of Figure~\ref{fig:temp_mass_distr} shows the effective temperature as a function of the orbital period. For comparison, also the values from \citet{Townsley+2009} and \citet{Pala+2017} are reported, together with the evolutionary tracks for a typical CV ($M_\mathrm{WD} = 0.8\,\mathrm{M}_\odot$, with an initial secondary mass of $M_2 = 0.65\,\mathrm{M}_\odot$ and an initial $P_\mathrm{orb} = 12\,$h, \citealt{Pala+2017}). 
With only two 150\,pc CVs having a $T_\mathrm{eff}$ at $P_\mathrm{orb} \gtrsim 180\,$min, the comparison with these tracks is only meaningful at short orbital periods. 
The $T_\mathrm{eff}$ distribution of the intrinsic population shows some fundamental discrepancies between the theory and the observations that have been already highlighted by \citet{Pala+2017} from the study of a large sample of CVs observed with the \textit{Hubble Space Telescope}. In particular, the systems at the period minimum are characterised by a large scatter in $T_\mathrm{eff}$ as the white dwarf temperatures due to compressional heating are very sensitive to
the white dwarf mass \citep{Townsley+2009}. The recent work by \cite{Diogo+2019} has shown that the mass distribution of CV white dwarfs is indeed the main reason behind this spread. Moreover, the systems below the period gap host hotter white dwarfs than suggested by the standard models (black track). This is also true in the case of the period bouncer track, which appears to be steeper than the theoretical predictions. As already discussed by \citet{Pala+2017}, these findings suggest that additional AML mechanisms are present (red track) besides pure GR in this period range. These additional AML mechanisms could also imply that period bounce CVs evolve faster than predicted by the models potentially explaining the lack of such evolved systems in the intrinsic population.

As shown in Figure~\ref{fig:temp_mass_distr} (left), the effective temperatures of systems with similar orbital periods but different distances (coloured against grey points) are fully consistent among each other, allowing us to rule out the presence of observational biases as possible explanation for the discrepancies between theory and observations. Our results, instead, highlights the need for a revision of the current models of compact binary evolution.

Finally, the masses of 14 CV white dwarfs in the 150~pc sample are available in the literature (Table~\ref{tab:cvwd_teff}, right panel of Figure~\ref{fig:temp_mass_distr}). The average mass results $\langle M_\mathrm{WD} \rangle = 0.83 \pm 0.17$, in perfect agreement with the measurement from \citet{Zorotovic}, $\langle M_\mathrm{WD} \rangle = 0.83 \pm 0.23\,\mathrm{M}_\odot$. As also shown by \citet{Zorotovic}, this result confirms that the higher mass of CV white dwarfs compared to that of single white dwarfs and their detached progenitors, $\langle M_\mathrm{WD} \rangle \simeq 0.6\,\mathrm{M}_\odot$ \citep{Koester1979,Liebert2005,Kepler2007}, cannot be related to an observational bias. 

\begin{figure}
\centering
\includegraphics[width=0.4\textwidth]{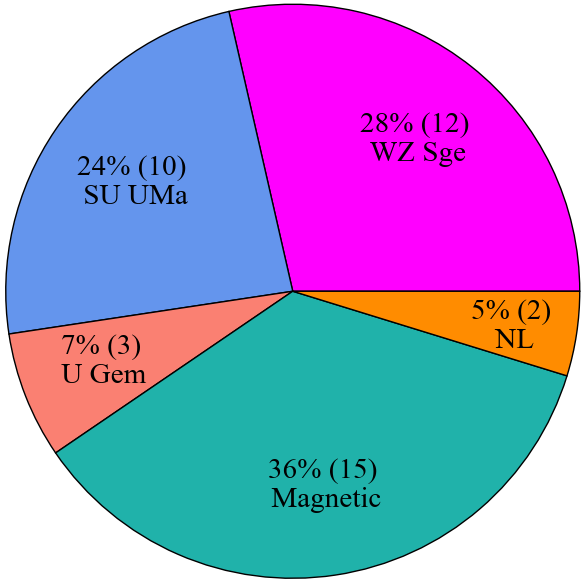}
\caption{CV subtype contribution to the 150~pc sample. More than one third of the observed CVs host a magnetic white dwarf, in clear contrast with the total absence of magnetic white dwarfs in the parent population of post common envelope binaries \citep{Liebert+2005,Liebert+2015}.}\label{fig:pie}
\end{figure}

The average mass of CV white dwarfs cannot be explained (i) invoking different parent populations for the present day CVs and the present day pre-CVs \citep[e.g.][]{Zorotovic} or (ii) assuming mass growth during nova cycles or through thermal time-scale mass transfer \citep{Wijnen+2015}.
Instead, the eCAML proposed by \citet{Schreiber+2016} could mitigate the discrepancy between the theory and the observations. The eCAML leads to merger of the two stellar components in systems hosting low mass white dwarfs, which would then disappear from the CV population thus naturally explaining the observed high average mass of CV white dwarfs. Nonetheless, the exact mechanism behind the additional CAML and the reason for its dependence on white dwarf mass are unclear. 

\subsection{Magnetic systems}
\citet{Pretorius+2013} derived the space density of magnetic CVs as $\rho_\mathrm{mCV}=1.3{+0.6\atop-0.4}\times10^{-6}\,\pc$. Comparing this value with the space density of non-magnetic CVs $\rho_0=4{+6\atop-2}\times 10^{-6}\,\pc$ from \citet{Pretorius+2012}, implies that about one third of CVs are magnetic. 

From the analysis of the \gaia parallaxes, we found that a large fraction, $(36 \pm 7)\,$per cent, of the CVs identified within 150~pc contain a magnetic white dwarf (Figure~\ref{fig:pie}). This corresponds to a space density\footnote{Note that the values reported in Table~\ref{tab:space_density_subtypes} cannot be corrected for incompleteness since this would require the knowledge of the contribution of each subtype to the population of CVs that still has to be discovered.} of $\rho_\mathrm{mCV}= 1.3{+0.3\atop-0.4}\times 10^{-6}\,\pc$ (Table~\ref{tab:space_density_subtypes}), consistent with the result by \citet{Pretorius+2013}.

Magnetic CVs are thought to follow a different evolutionary path compared to non-magnetic system. Owing to the coupling with the magnetic field of the white dwarf, MB is reduced or even completely suppressed in the system with the strongest magnetic fields \citep{Li+1994}. Consequently, IPs and polars evolve slower than non-magnetic CVs, thus explaining the high fraction of observed magnetic CVs \citep{Araujo-Betancor+2005a}. 

However, this high incidence of magnetism in CVs is not reflected in the fraction of magnetic white dwarfs observed in their parent population, i.e. the post common envelope binaries (PCEBs), with no confirmed magnetic white dwarf detected in any of these detached binaries \citep{Liebert+2005,Liebert+2015}. 
Different scenarios have been proposed to explain the observed fraction of magnetic CVs, such as Ap and Bp progenitors which preserve a fossil magnetic field while becoming white dwarfs \citep{Angel+1981} or interaction during the common envelope phase \citep{Tout+2008}. However, they have been unable to explain the lack of magnetic white dwarfs in the population of PCEB and, consequently, the origin of magnetic CVs remains unclear.    

\begin{table}
\centering
\caption{Space densities for different CV sub-types computed considering the volume enclosed within $d_\mathrm{lim} = 150$\,pc assuming a scale height $h = 280\,$pc.}\label{tab:space_density_subtypes}
\begin{tabular}{lccc}
\toprule
Subtype & $N_\mathrm{CV} $ & $N_\mathrm{CV}/N_\mathrm{tot}$ & $\rho_0~[\times 10^{-6}$~\pc] \\
\midrule
WZ\,Sge       &  12 & ($28\pm7$)~\% & $1.0 \pm 0.3$ \\
  \addlinespace[0.2cm]
SU\,UMa       &  10 & ($24\pm7$)~\% & $0.8{+0.2\atop-0.4}$ \\
  \addlinespace[0.2cm]
U\,Gem        &  3 & ($7\pm4$)~\%   &  $0.2{+0.2\atop-0.1}$ \\
  \addlinespace[0.2cm]
NL            & 2 & ($5\pm3$)~\%    &  $0.15{+0.2\atop-0.3}$ \\
  \addlinespace[0.2cm]
Magnetic      &  15 & ($36\pm7$)~\% &  $1.3{+0.3\atop-0.4}$ \\

\bottomrule  
\end{tabular}
\end{table}

\subsection{Systems with an evolved donor}\label{subsec:evolved_donors}
The stability of the mass transfer process requires a mass ratio $q = M_2/M_\mathrm{WD} \lesssim 1$ (where $M_2$ is the mass of the secondary star, \citealt{Frank+2002}). However, it has been shown that CVs hosting massive donors ($M_2 \simeq 1.5\,\mathrm{M}_\odot$) could survive a phase of thermal-time-scale mass transfer during which the accreted material burns steadily onto the white dwarf \citep{Schenker+2002}. During the thermal-time-scale mass transfer,  the donor is stripped of its envelope and the surviving system is a normal CV in which the white dwarf accretes from the remnant core of its companion, rich of CNO processed material. These CVs are predicted to make up for $\simeq 30\,$per cent of the Galactic CV population \citep{Schenker+2002}. 

These systems can be easily recognised in the ultraviolet from their enhanced \ion{N}{v}/\ion{C}{iv} line flux ratios compared to those of CVs that have formed through the standard channel and the observed fraction of CVs with an evolved donor has been found to be $\simeq 15\,$per cent \citep{Boris+2003}. 
In the 150~pc CV sample, we identified only two of such CVs: AE\,Aqr \citep{Jameson+1980} and V2301\,Oph, \citep{Schmidt+2001}, corresponding to a fraction of $(5 \pm 3)\,$per cent. The higher fraction found by \citet{Boris+2003} can be explained by an observational bias, these systems host more massive (and hence larger) and brighter donors compared to those of normal CVs and therefore are easily detected even at large distances. Instead, the theoretically predicted fraction by \citet{Schenker+2002}, 30\,per cent, is much higher compared to the observed fraction of CV with evolved donors in the 150~pc sample. 

CVs with evolved donors represent an evolutionary link between CVs and the more compact AM\,CVn stars. The formation of the latter is still poorly understood but three different pathways have been proposed \citep{Nelemans+2005}. In the first scenario, the systems are formed from a double white dwarf binary in which the less massive star is brought to contact with its Roche lobe by the orbital shrinkage due to GWR. Alternatively, the progenitors could be a binary containing a white dwarf plus a non-degenerate, helium core burning star. Finally, AM\,CVn could descend from CVs with an evolved donor. Hosting massive secondary stars, these systems are able to evolve to $P_\mathrm{orb}$ much shorter than the period minimum. Owing to the mass transfer process, the secondary is progressively eroded, till only its Helium-rich core is left and an AM\,CVn star is born.
The fraction of CVs with an evolved donor that we derived could provide an upper limit on the number of AM\,CVn that are expected to form through this channel, yielding a valuable observational test for the contribution of the different formation channels to the overall population of these compact systems. 

  \begin{table*}
\caption{150~pc CVs with a common proper motion companion.}\label{tab:tripletes}
\setlength{\tabcolsep}{0.02cm}
\begin{adjustwidth}{-2.5em}{-2em}
\begin{tabular}{@{}llcccccccccc@{}}
 \toprule
 System         & \multicolumn{1}{c}{\gaia DR2 ID} & $P_\mathrm{orb}$ & Type & $\varpi$ & $\mu_\mathrm{RA}$ & $\mu_\mathrm{Dec}$ & $\Delta$ & Angular & \multicolumn{2}{c}{$D$} & $-E_\mathrm{bind}$  \\ 
                      &                                                     &    &         &   &   &     & & separation &  & \\
                     &                                                     &   (min)                    &       &  (mas)    &   (mas~yr$^{-1}$)               &   (mas~yr$^{-1}$)       &           &   & (pc) & (AU) & (erg) \\ 
\midrule 
V379\,Tel                              & 6658737220627065984 & 101.03 & AM & 7.65(7) & $-60.8(1)$ & $-17.75(9)$ &  \multirow{2}{*}{2.9} & \multirow{2}{*}{30.8\arcsec} & \multirow{2}{*}{$2.3 \pm 1.5$} &\multirow{2}{*}{$(5 \pm 3) \times 10^{5}$} & \multirow{2}{*}{$(1.5 \pm 1.0) \times 10^{40}$}  \\ 
                                               & 6658737388128701184 &    &   M2V   & 7.52(5) & $-60.64(8)$ & $-18.44(6)$ & & \\  

\addlinespace[0.3cm]
\gaia\,J154008.28--392917.6 & 6008982469163902464 & -- & UGWZ & 7.49(11) & $64.8(2)$ & $1.6(2)$   &  \multirow{2}{*}{2.4} & \multirow{2}{*}{1.1\degree} &\multirow{2}{*}{$3 \pm 13$} &\multirow{2}{*}{$(6 \pm 26) \times 10^{5}$} & \multirow{2}{*}{$(2 \pm 8) \times 10^{40}$}\\ 
                         & 6002961479076981632 &  & WD  & 7.6(8) & $52(2)$ & $4(1)$ &  & \\ 

\addlinespace[0.3cm]
\gaia\,J051903.99+630340.4 & 285957277597658240 & 126: & UGSU & 8.59(4) & $-13.07(5)$ & $-45.30(5)$   &  \multirow{2}{*}{11} & \multirow{2}{*}{6.8\arcsec} &\multirow{2}{*}{$2.2 \pm 0.7$} &\multirow{2}{*}{$(4.6 \pm 1.4) \times 10^{5}$} & \multirow{2}{*}{$(3.7 \pm 1.1) \times 10^{40}$}\\ 
TYC\,4084-172-1                     & 285957277597658368 &    & G3V & 8.43(3) & $-11.19(3)$ & $-44.38(4)$ &  & \\

\bottomrule
\end{tabular}
\end{adjustwidth}
\begin{tablenotes}
\item \textbf{Notes.}
$-\mathit{E}_\mathrm{bind}$ has been computed assuming $M_3 = 0.4\,\mathrm{M}_\odot$, $M_3 = 0.6\,\mathrm{M}_\odot$ and 
$M_3 = 1\,\mathrm{M}_\odot$ for V379\,Tel, \gaia\,J154008.28--392917.6 and \gaia\,J051903.99+630340.4, respectively.
\end{tablenotes}
\end{table*}

\begin{figure}
\centering
\includegraphics[width=0.48\textwidth]{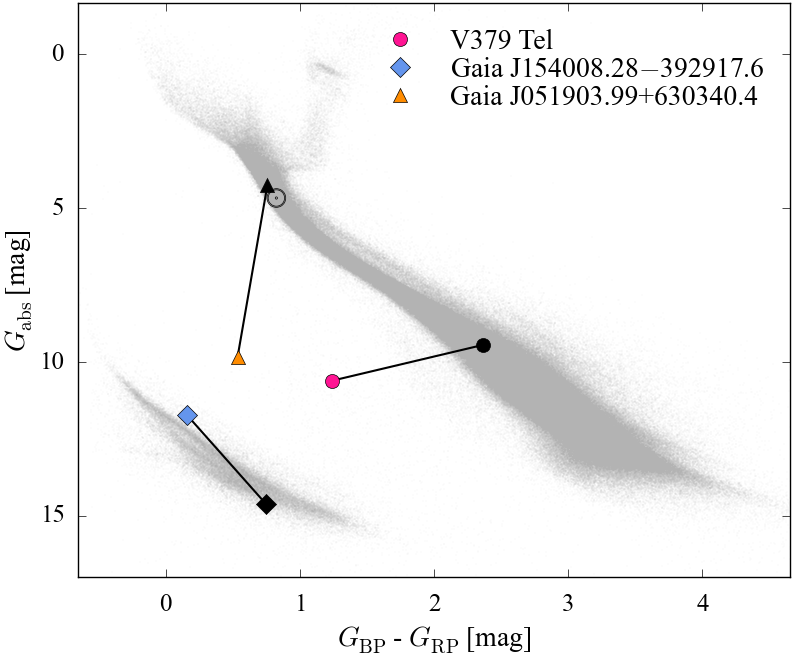}
\caption{Hertzsprung-Russell diagram of the \gaia sources within 150\,pc (grey) showing, as a reference, the position occupied by the Sun \citep{Casagrande+2018} and by V379\,Tel (pink), \gaia\,J154008.28--392917.6 (blue) and \gaia\,J051903.99+630340.4 (orange) and their common proper motion companions (black), which result being a M-dwarf, a white dwarf (WD) and a G-type star, respectively.}\label{fig:hr3}
\end{figure}

\section{Tertiary companions}\label{sec:triplets}
The formation and the evolution of a binary can be influenced by the presence of a nearby third body orbiting the system. For example, the third body can give rise to Lidov-Kozai cycles that can induce periodical variation in the eccentricity and the inclination of the inner binary with respect to the orbital plane of the external body (see \citealt{Naoz2016} for a review). This gravitational interaction can hence affect the mass transfer process and potentially even lead to mergers \citep[e.g.][]{Toonen+2016}. In order to fully constrain the formation and evolution of CVs, it is important to assess the multiplicity fractions of these systems.

In the past, the most successful method to identify wide orbit companions was the identification of period modulations in the long term light curves of the system and several studies have suggested that some CVs could be part of triple systems. Few examples are: VY\,Scl \citep{Martinez-Pais+2000}, DP\,Leo \citep{Beuermann+2011}, FS\,Aur, \citep{Chavez+2012} and LX\,Ser \citep{Li+2017}, which have been suggested to host circumbinary sub-stellar objects or giant planets. However, the observed modulations could also been explained by the magnetic activity of the donor star and, owing to the lack of a direct detection of the third bodies, it is difficult to disentangle the two scenarios. 

The accurate parallaxes and proper motions delivered by the \gaia space mission in its DR2 offer the first opportunity to carry out a systematic search for third components on wide orbits to the 150~pc CVs by searching for common proper motion companions.

In order to identify these objects, we performed a cone search of 3\,pc radius around each of the 150~pc CVs, and selected those objects for which
\begin{equation}
\Delta = \sqrt{\left( \frac{\Delta \mu_\mathrm{RA}}{\sigma_{\mu_\mathrm{RA}}}\right)^2 +  \left( \frac{\Delta \mu_\mathrm{Dec}}{\sigma_{\mu_\mathrm{Dec}}}\right)^2} < 3
\end{equation}
where $\Delta \mu$ are the differences in the proper motion components and $\sigma_{\mu}$ are the quadrature sum of the corresponding $3\sigma$ uncertainties.

We find that two CVs, V379\,Tel and \gaia\,J154008.28--392917.6, satisfy this condition. Using the \gaia coordinates and parallaxes, we computed the separation between the inner binary and the third body which resulted $D \simeq 2.5\,$pc (Table~\ref{tab:tripletes}). Although it did not satisfy the previous condition ($\Delta \simeq  11$), we cannot ignore that \gaia\,J051903.99+630340.4 has a nearby companion (TYC\,4084-172-1) located at a similar distance ($\simeq 2.2\,$pc) and therefore we also include this object in the following discussion.

Given the relatively large orbital separations, we computed the gravitational binding energy:
\begin{equation}
-E_\mathrm{bind} = G\frac{(M_\mathrm{WD}+M_2)M_3}{D}
\end{equation}
of each triplet assuming that the inner binary is a typical short period CV (as the three systems listed in Table~\ref{tab:tripletes} are all located below the period gap), with a white dwarf of mass $M_\mathrm{WD} = 0.83\,\mathrm{M}_\odot$ and a donor of mass $M_2 = 0.14\,\mathrm{M}_\odot$, typical of a CV below the period gap \citep{Knigge+2011}. We assumed that the third body is on a circular orbit around the inner binary and we estimated their possible masses accordingly to their position on the Hertzsprung-Russell diagram (Figure~\ref{fig:hr3}). The third companions result being a M-dwarf of spectral type $\simeq\,$M2V (as we estimated using as a reference the table\footnote{\url{http://www.pas.rochester.edu/\~emamajek/EEM\_dwarf\_UBVIJHK\_colors\_Teff.txt}} from \citealt{Pecaut+2013}), a white dwarf (WD) and a G-type star for V379\,Tel, \gaia\,J154008.28--392917.6 and \gaia\,J051903.99+630340.4, respectively. We therefore assumed:
\begin{itemize}
\item $M_3 = 0.4\,\mathrm{M}_\odot$, corresponding to the mass of a M2V star \citep{Pecaut+2013}, in the case of the tertiary companion of V379\,Tel;
\item $M_3 = 0.6\,\mathrm{M}_\odot$, corresponding to the average mass of single white dwarfs \citep{Liebert2005}, in the case of the tertiary companion of \gaia\,J154008.28--392917.6;
\item $M_3 = 1\,\mathrm{M}_\odot$ in the case of TYC\,4084-172-1, given its spectral type G3V and surface gravity of $\log(g) = 4.2(2)$ (\citealt{Frasca+2018}, see also Figure~\ref{fig:tyc}).
\end{itemize}
The binding energy we derived are listed in Table~\ref{tab:tripletes}. The related uncertainties have been computed accounting for those on the \gaia coordinates and parallaxes of each pair of objects, and are dominated by the latter. At this stage, their large values combined with the unknown uncertainties related to the theoretical limit for stellar binding energies, $-E_\mathrm{bind} \approx 1 \times 10^{41}$ \citep{Burgasser+2007}, make it impossible to unambiguously establish whether these CVs are part of hierarchical systems and additional observational efforts are required in order to finally assess their hierarchical structure.

\section{Conclusions}
Making use of the accurate astrometry delivered by the ESA \gaia space mission in its DR2, we carry out the first detailed study of the volume-limited sample of 42 CVs located within 150~pc. Combining the \gaia data with the photometric and spectroscopic observations from SDSS and CRTS, we estimate the sample to be $(77 \pm 10)\,$per cent complete. This is mainly dictated by the efficiency of the discovery methods employed in detecting new CVs, which are biased towards systems accreting at intermediate mass rates, that can be easily detected in the X-rays or thanks to their dwarf nova outbursts.

Assuming $h=280\,$pc as a typical scale heigh for the Galactic CV population, we estimate the CV space density, which results $\rho_0=(4.8^{+0.6}_{-0.8}) \times10^{-6}\,\pc$. Thanks to the exquisite \gaia data, we reduce the uncertainty on $\rho_0$ by a factor of ten compared to the pre-\gaia estimates.
The uncertainties we derive take well into account for the estimated completeness of the sample and for possible different values of the scale height in the range $100 - 500\,$pc. Nonetheless, given that the 150~pc CV sample is dominated by short period systems representing the old component of the Galactic CV population, it is reasonable to assume that the larger values of $h$ are likely the most realistic.

The advent of the \gaia space mission provides also the unique opportunity to study the intrinsic properties of the Galactic CV population and to constrain the models describing the formation and evolution of these systems. We find that the observed space density is significantly lower than predicted by the current available models of CV evolution. Moreover, the fractions of CVs above (17\,per cent) and below (83\,per cent) the period gap are in clear disagreement with the theoretical predictions (1\,per cent and 99\,per cent, respectively).
Both discrepancies can be resolved by the recently proposed eCAML model in which CVs hosting low-mass white dwarfs merge owing to frictional AML arising from nova explosions. Consequently, the  Galactic CV population would be composed by a lower absolute number of systems that would naturally imply a lower space density. Moreover, the fractions of CVs predicted by the eCAML model results 15\,per cent and 85\,per cent above and below the period gap respectively, in better agreement with the observed value. The disappearance of the CVs hosting low-mass white dwarfs would also be consistent with the average masses of the CVs white dwarfs in the 150\,pc sample, $\langle M_\mathrm{WD} \rangle = 0.83 \pm 0.17\,\mathrm{M}_\odot$, being higher than the masses of their detached progenitors (i.e. PCEB, $\langle M_\mathrm{WD} \rangle \simeq 0.6\,\mathrm{M}_\odot$, \citealt{Zorotovic}).
The need to include additional mechanisms of AML in the models is also supported by the observed effective temperatures and, consequently, mass accretion rates of the 150\,pc CVs, which, below the period gap, are found to be accreting at higher rates than theoretically predicted.
However, the observed fraction of period bounce CVs, 7-14\,per cent is much lower than current models (including eCAML) predicts, $40-80$\,per cent and it is possible that the time scales required for a CV to evolve to the period minimum are much longer than theoretically predicted

Studying the composition of the 150\,pc CV sample, we identify a large fraction of magnetic CVs, 36\,per cent. 
This finding is particularly intriguing given that no confirmed magnetic white dwarf is known among the PCEBs. All the models proposed to explain the observed fraction of magnetic CVs predict also the existence of magnetic white dwarfs in PCEBs and, consequently, this high incidence of magnetism among CV white dwarfs remains unclear.

We also show that the fraction of CVs hosting nuclear evolved donors is $\simeq 5$ per cent, lower than the pre-\gaia observational estimate, $\simeq 15$ per cent. Most likely, this difference arises from an observational bias since these systems are brighter that normal CVs and are easily detected even at large distances. Moreover, the observed fraction of CVs with evolved donors is significantly lower than predicted by the theory (30\,per cent).

Finally, we find that three CVs have a common proper motion companion. However, the lack of accurate system parameters does not allow to draw definite conclusion on whether they form hierarchical triple systems and additional observations are required to finally establish whether these common proper motion companion pairs are gravitationally bound.

\section*{Acknowledgements}
This work has made use of data from the European Space Agency (ESA) mission {\it Gaia} (\url{https://www.cosmos.esa.int/gaia}), processed by the {\it Gaia} Data Processing and Analysis Consortium (DPAC, \url{https://www.cosmos.esa.int/web/gaia/dpac/consortium}). Funding for the DPAC has been provided by national institutions, in particular the institutions participating in the {\it Gaia} Multilateral Agreement.

The research leading to these results has received funding from the European Research Council under the European Union's Seventh Framework Programme (FP/2007--2013) / ERC Grant Agreement n. 320964 (WDTracer). 

Based on observations made with ESO Telescopes at Paranal Observatory under programme ID 0101.C-0646(A).

Based on observations obtained under programme ID SO2018B-015 at the Southern Astrophysical Research (SOAR) telescope, which is a joint project of the Minist\'{e}rio da Ci\^{e}ncia, Tecnologia, Inova\c{c}\~{o}es e Comunica\c{c}\~{o}es (MCTIC) do Brasil, the U.S. National Optical Astronomy Observatory (NOAO), the University of North Carolina at Chapel Hill (UNC), and Michigan State University (MSU).

The research leading to these results has received funding from the European Research Council under the European Union's Horizon 2020 research and innovation programme n. 677706 (WD3D).

The work presented in this article made large use of \textsc{TOPCAT} and \textsc{STILTS} Table/VOTable Processing Software \citep{Topcat}.

Funding for the SDSS and SDSS-II has been provided by the Alfred P. Sloan Foundation, the Participating Institutions, the National Science Foundation, the U.S. Department of Energy, the National Aeronautics and Space Administration, the Japanese Monbukagakusho, the Max Planck Society, and the Higher Education Funding Council for England. The SDSS Web Site is http://www.sdss.org/.

The SDSS is managed by the Astrophysical Research Consortium for the Participating Institutions. The Participating Institutions are the American Museum of Natural History, Astrophysical Institute Potsdam, University of Basel, University of Cambridge, Case Western Reserve University, University of Chicago, Drexel University, Fermilab, the Institute for Advanced Study, the Japan Participation Group, Johns Hopkins University, the Joint Institute for Nuclear Astrophysics, the Kavli Institute for Particle Astrophysics and Cosmology, the Korean Scientist Group, the Chinese Academy of Sciences (LAMOST), Los Alamos National Laboratory, the Max-Planck-Institute for Astronomy (MPIA), the Max-Planck-Institute for Astrophysics (MPA), New Mexico State University, Ohio State University, University of Pittsburgh, University of Portsmouth, Princeton University, the United States Naval Observatory, and the University of Washington.

Funding for SDSS-III has been provided by the Alfred P. Sloan Foundation, the Participating Institutions, the National Science Foundation, and the U.S. Department of Energy Office of Science. The SDSS-III web site is http://www.sdss3.org/.

SDSS-III is managed by the Astrophysical Research Consortium for the Participating Institutions of the SDSS-III Collaboration including the University of Arizona, the Brazilian Participation Group, Brookhaven National Laboratory, Carnegie Mellon University, University of Florida, the French Participation Group, the German Participation Group, Harvard University, the Instituto de Astrofisica de Canarias, the Michigan State/Notre Dame/JINA Participation Group, Johns Hopkins University, Lawrence Berkeley National Laboratory, Max Planck Institute for Astrophysics, Max Planck Institute for Extraterrestrial Physics, New Mexico State University, New York University, Ohio State University, Pennsylvania State University, University of Portsmouth, Princeton University, the Spanish Participation Group, University of Tokyo, University of Utah, Vanderbilt University, University of Virginia, University of Washington, and Yale University.

A.A. acknowledges generous supports from Naresuan University
M.R.S. thanks for support from Fondecyt (grant 1181404).
O.T. was supported by a Leverhulme Trust Research Project Grant.
B.T.G. and O.T. were supported by the UK STFC grant ST/P000495.




\bibliographystyle{mnras}
\bibliography{main} 




\appendix

\section{Flaring red dwarfs}\label{ap:frd}
\begin{figure}
\includegraphics[width=0.48\textwidth]{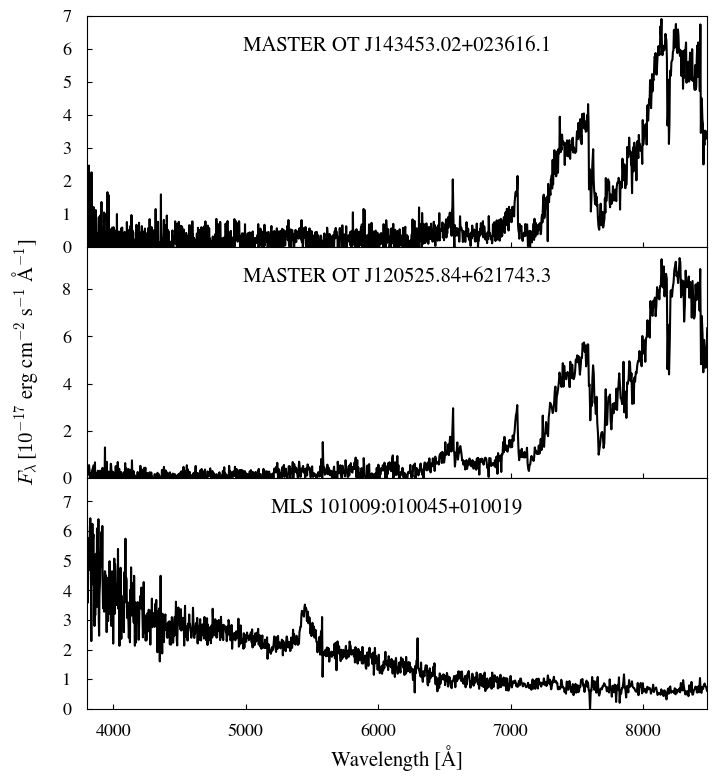}
\caption{Sample systems reported as CV candidates in the literature. The SDSS spectra reveil that they are actually flaring red dwarfs (top and middle panels) and a quasar (bottom panel).}\label{fig:master_mls}
\end{figure}

12 objects in our sample are located on the main sequence or very close it (empty and yellow diamonds in Figure~\ref{fig:hr_all}).
These objects are often classified as CV candidates because they have shown some transient phenomena that has been interpreted as a likely dwarf nova outburst. However, these systems are much redder than the typical CVs and it is therefore more likely that they are actually flaring red dwarfs.
This is the case, for example, of MASTER\,OT\,J143453.02+023616.1 and MASTER\,OT\,J120525.84+621743.3: their SDSS spectra confirm their red dwarf nature (top and middle panel of Figure~\ref{fig:master_mls}).
The remaining 10 systems have similar colors and we conclude that they are also flaring red dwarfs.

\begin{figure*}
\includegraphics[width=\textwidth]{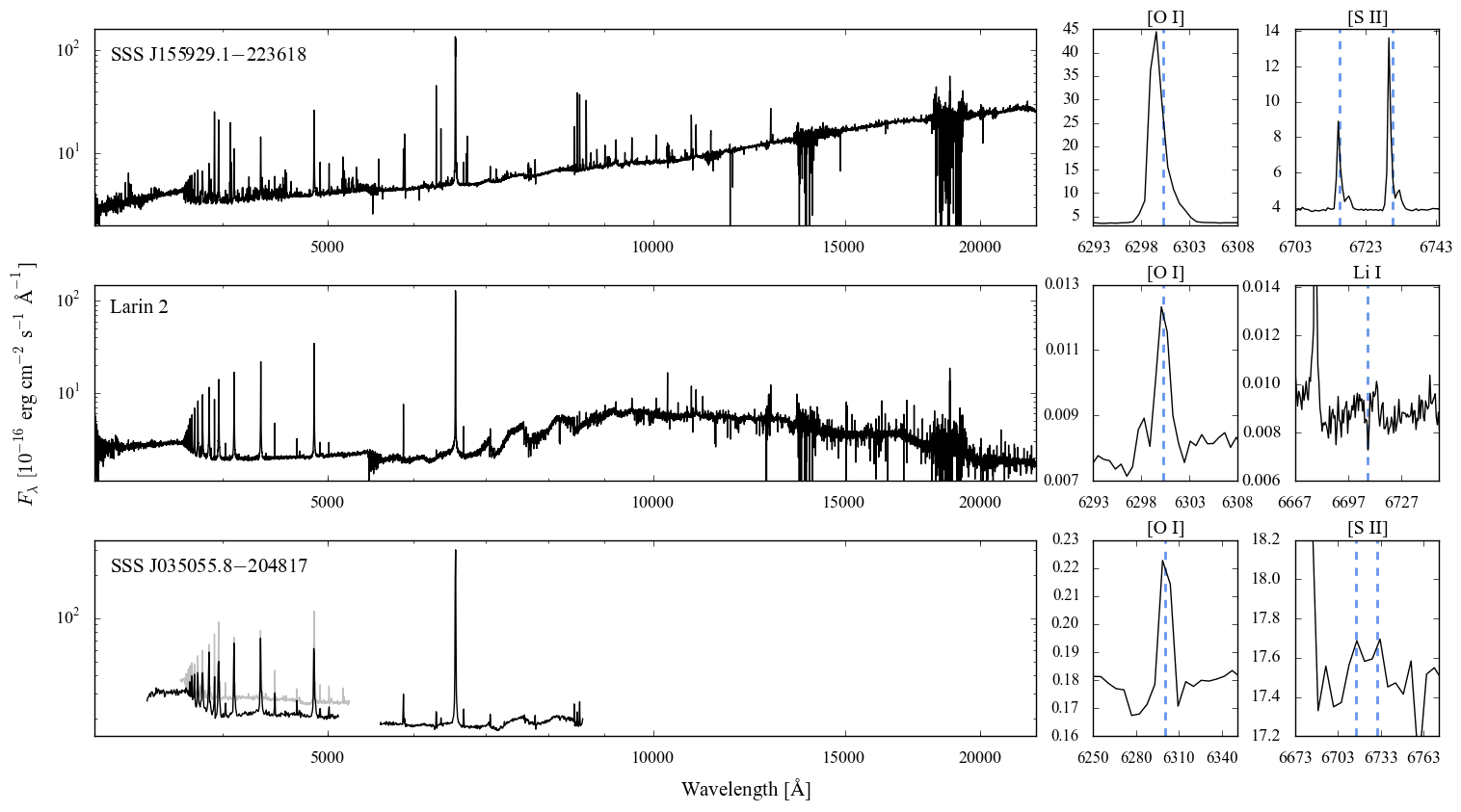}
\caption{VLT/X-shooter (top and middle panels), SOAR (grey, bottom panel) and WHT/ISIS (black, bottom panel) spectra of three YSOs  that have been mistakenly classified as CVs \citep{Watson+2006,Larin+2018}. In all the three systems, the metal emission lines are much narrower than the hydrogen lines, suggesting that the first arise from a hot corona close to the stellar surface while the seconds originate far-out in the magnetosphere. The panels on the left show a zoom in the regions in which \ion{Li}{i} and/or the forbidden lines of [\ion{O}{i}] (6300\,\AA) and [\ion{S}{ii}] (6730\,\AA) are detected, which are all characteristic of YSOs.}\label{fig:yso}
\end{figure*}

\begin{figure*}
\includegraphics[width=\textwidth]{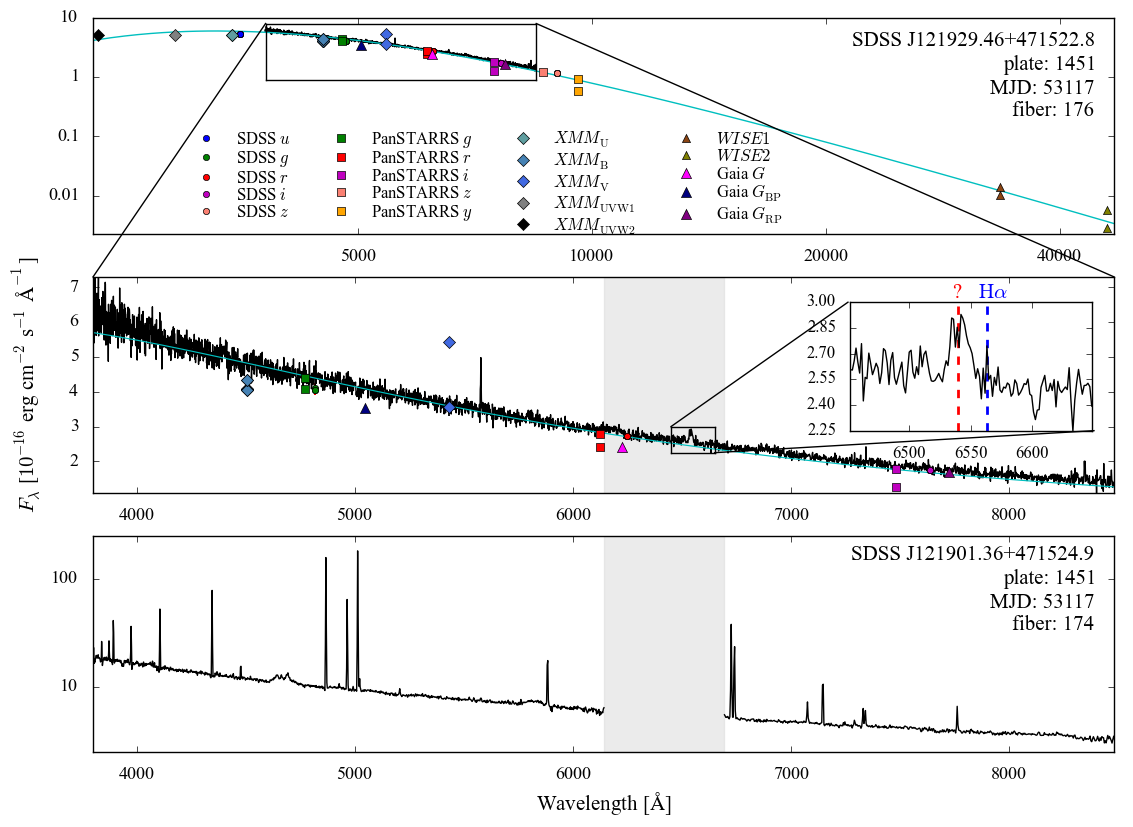}
\caption{Photometric SED (top) and SDSS spectrum (middle) of SDSS1219 along with a black body ot $T_\mathrm{eff} = 8000\,$K. On the bottom panel is shown the spectrum of a galaxy observed by a nearby fiber on the same SDSS plate. This galaxy saturated the SDSS detector in the wavelength region (grey band, also highlighted in the middle panel) in which the emission lines is observed in SDSS1219 and it most likely the origin of this anomalous feature.}\label{fig:sdss1219}
\end{figure*}

\begin{figure*}
\includegraphics[width=\textwidth]{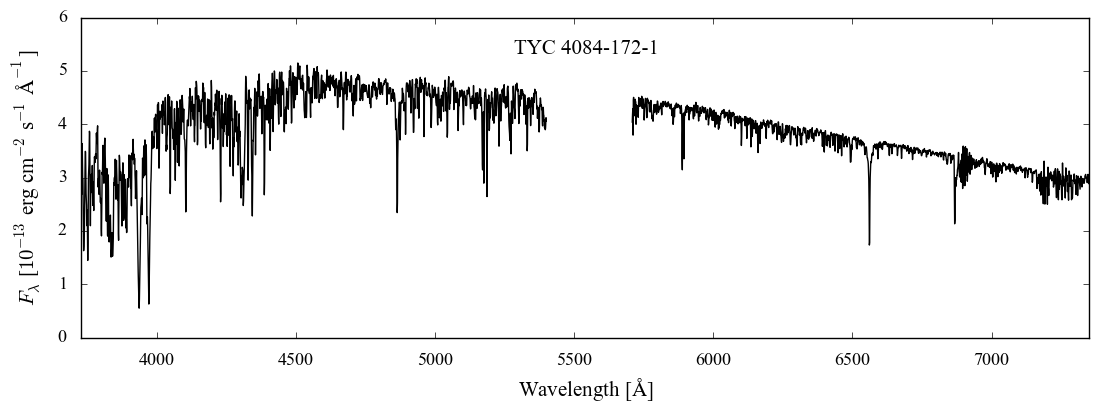}
\caption{WHT/ISIS spectrum of TYC\,4084-172-1, a G3V star \citep{Frasca+2018}, located 2.2\,pc away from \gaia\,J051903.99+630340.4}.\label{fig:tyc}
\end{figure*}

\section{Young stellar objects}\label{ap:yso}
Three CV candidates, SSS\,J155929.1-223618 \citep{Watson+2006}, Larin\,2 \citep{Larin+2018,Denisenko+2018} and SSS\,J035055.8-204817 show particularly red colours but have also a \textit{Galex} detection. \citet{Larin+2018} suggested that Larin\,2 could be a long period magnetic CVs with a large infrared contribution from the secondary. Later on \citet{Denisenko2018} argued that wind-driven accretion is going on in this system, where the wind circularisation radius is smaller than the Roche lobe radius. 

We acquired VLT/X-shooter and William Herschel Telescope/Intermediate-dispersion Spectrograph and Imaging System (WHT/ISIS) spectroscopic for these systems, as well as one SOAR spectrum for SSS\,J035055.8-204817, (Figure~\ref{fig:yso}), where we detected the presence of the lithium absorption line at $6707\,$\AA. Lithium is expected to be depleted once the core is hot enough for it to be burnt, and since low-mass stars at this stage are fully convective, its presence in the stellar photosphere suggests youth \citep{Soderblom+2014}. Moreover, the forbidden lines of [\ion{O}{i}] (5577\,\AA\, and 6300\,\AA) and [\ion{S}{ii}] (6730\,\AA) are commonly observed in young stellar objects (YSOs, \citealt{Fang+2018}) but not in CVs. Finally, the metal emission lines are much narrower than the hydrogen lines and these different line widths suggest that the first arise from a hot corona close to the stellar surface while the seconds originate far-out in the magnetosphere \citep{Hamann+1992}. We therefore concluded that both systems are likely YSOs (brown diamonds in Figure~\ref{fig:hr_all}).

Recently, \citet{Denisenko+2018} identified another CV candidate within 150 pc, DDE\,158, owing to its colours and variability properties similar to those of Larin\,2. Similarly to SSS\,J155929.1-223618, Larin\,2 and SSS\,J035055.8-204817, the SOAR spectrum of DDE\,158 presented in the vsnet-alert \#23884 is characterised by metal emission lines narrower than the hydrogen ones, suggesting that they do not originates in a typical CV accretion disc. The non-detection of the forbidden  [\ion{O}{i}] (5577\,\AA) line does not allow do rule out a YSO (has suggested in the vsnet-alert \#23884) since several YSOs in which this feature has not been detected (even in medium resultion spectra, see e.g. \citealt{Natta+2014, Nisini+2018,Banzatti+2019}) are known. Finally, the red portion of the spectrum resembles the emission of a M-dwarf star. If DDE\,158 was a CV, a secondary star of such spectral type would imply the presence of a white dwarf as hot as $\gtrsim 15\,000\,$K. Given the distance as derived from the \gaia parallax ($\simeq 103\,$pc), such white dwarf would contribute significantly to the overall system emission, causing the spectrum to be much bluer than actually observed. We can thus conclude that also DDE\,158 is likely another YSO and therefore we do not include it in our sample.

\section{Other non-CV systems}\label{ap:ncvs}
Three systems listed in the literature as CVs, SDSS\,J121929.46+471522.8 \citep{Szkody+2006}, NSV\,15401 \citep{Downes+2001} and Gaia14abg \citep{Rixon+2014}, are actually single white dwarfs.

The SDSS spectrum of SDSS\,J121929.46+471522.8 shows a blue continuum and a weak emission line at $\lambda = 6539.4\,$\AA\, but no absorption and/or emission lines typical of a CV (Figure~\ref{fig:sdss1219}). This is likely a cold ($T_\mathrm{eff} \simeq 8000\,K$) white dwarf whose atmosphere is dominated by helium, which explain the absence of absorption features. The origin of the emission is however unclear. Its wavelength seems to suggest that it arises from a \ion{O}{ii} transition. However, this is quite unlikely because (i) the formation of \ion{O}{ii} lines require much higher temperatures than the one we estimated for SDSS1219 and (ii) many other stronger \ion{O}{ii} lines should be detected in the spectrum. Contamination from a nearby fibre in the SDSS plate, centred onto a source showing strong emission in this wavelength is, more likely, the origin of the observed emission in SDSS1219. 

The spectrum of NSV\,15401 (aka Lan\,159) presented by \citeauthor{Lepine+2011} (\citeyear{Lepine+2011}, see their figure~2) clearly show that this is a single white dwarf while Gaia14abg was among the first objects identified by the \gaia alerts and was affected by a cross-matching problem. 

We also discarded five detached binaries (Ret1 REF, BPM\,18764 REF), one pre-polar (WX\,LMi), and MLS\,101009:010045+010019, whose SDSS spectrum shows that this system is actually a quasar (bottom panel of Figure~\ref{fig:master_mls}).

\section{New CVs from \gaia}\label{ap:new_cvs}
We observed \gaia\,J051903.99+630340.4 with the Intermediate-dispersion Spectrograph and Imaging System (ISIS) spectrograph mounted on the William Herschel Telescope in La Palma (Spain). We used a 1.2\arcsec slit combined with the R600B and R600R gratings, centred at 4540~\AA\, and 6561~\AA\, and providing a nominal spectral resolution of $R \simeq 2000$ and $R \simeq 4000$, respectively.
\gaia\,J051903.99+630340.4 shows a spectrum typical of a SU\,UMa star, dominated by a blue continuum and strong double-peaked Balmer and \ion{He}{i} emission lines (top panel of Figure~\ref{fig:new_gaia_cvs}). Although this class of CVs is characterised by relatively short outburst recurrence times (of the order of months up to one year), it is possible that the dwarf nova outbursts of \gaia\,J051903.99+630340.4 have been missed given the presence of the much brighter nearby companion (TYC\,4084-172-1, $G = 9.6\,$mag, see Figure~\ref{fig:tyc} and Section~\ref{sec:triplets}).
With the ISIS spectrograph, we also obtained phase resolved spectroscopic observations. From a fit to the position of the H$\alpha$ emission, we estimated a $P_\mathrm{orb} \simeq 126 \,$min, consistent with the SU\,UMa classification.

We performed a spectroscopic follow-up of \gaia\,J154008.28--392917.6 using the Goodman spectrograph \citep{Clemens+2004} mounted on the Southern Astrophysical Research (SOAR) telescope in Cerro Pach\'on (Chile). We used a 1\arcsec\, slit and the red camera to acquired eight spectra of 300\,s exposure each, using a 930 line/mm grating covering the wavelength range $3650\,$\AA$ - 5200\,$\AA. 
The average spectrum is shown in the bottom panel of Figure~\ref{fig:new_gaia_cvs}. \gaia\,J154008.28--392917.6 resembles a typical low accreting system, with the signature of both the white dwarf (pressure broadened Balmer absorption lines) and the accretion disc (double-peaked Balmer emission lines) clearly detected in its spectrum. Given its spectral similarities with other CVs at the period minimum (see e.g. EZ\,Lyn in Figure~\ref{fig:sdss_spectra}), we classified \gaia\,J154008.28--392917.6 as WZ\,Sge-type CV likely located close (or even having already evolved through) the period minimum.

\clearpage
\onecolumn
\setlength{\tabcolsep}{0.1cm}
\begin{longtable}{lcccccc}
\caption{CVs with unreliable \textit{Gaia} parallaxes (Section~\ref{sec:aen} \& \ref{sec:spurious}) and objects within 150~pc that have been mistakenly identified as CVs in the literature. The flags are as follows: FRD, likely flaring red dwarf; SWD, single white dwarf; DB, detached binary; YSO, young stellar object; Q, quasar; AEN, astrometric excess noise greater than 2; SGD, spurious \gaia detection.}\label{tab:discarded}\\
 \toprule
 System         & $\alpha$ & $\delta$ & $\varpi$ & $\sigma_\varpi$ & Astrometric  & Flag \\
                     &               &               &  (mas)    &   (mas)               &  excess noise &       \\ \midrule
 \endfirsthead
\multicolumn{7}{c}%
{{\tablename\ \thetable{} -- continued from previous page}} \\
\toprule
 System         & $\alpha$ & $\delta$ & $\varpi$ & $\sigma_\varpi$ & Astrometric  & Flag \\
                     &               &               &  (mas)    &   (mas)               &  excess noise &       \\ \midrule
\endhead
\bottomrule
\endfoot
\bottomrule
\endlastfoot
MASTER\,OT\,J015119.13--643046.6 & 01:51:19.39	& --64:30:45.18 & 25.1 &	0.3 & 1.15 & FRD \\
MASTER\,OT\,J031121.54--601851.0 & 03:11:21.43 & --60:18:50.24 & 19.07 & 0.13 & 0.96 & AEN \\
OGLE-BLG-DN-0128                           & 17:47:29.66 & --34:42:44.55 & 18 & 1 & 2.12 & AEN \\
N\,SMC\,2012 & 00:32:55.06 & --74:20:19.7 & 14.60 & 0.09 & 0.50 & SGD \\
SDSS\,J121929.46+471522.8			& 12:19:29.32 & +47:15:22.89 & 14.3 & 0.1 & 0.46 & SWD \\ 
CSS\,131106:032129+180827			& 03:21:28.62 & +18:08:27.05 & 12.8 & 0.5 & 1.64 & FRD \\
MASTER\,OT\,J020836.79--104018.8	& 02:08:36.74 & --10:40:18.79 & 12.2 & 0.4 & 1.58 & FRD\\
MASTER\,OT\,J194753.58--475722.9 & 19:47:53.60 & --47:57:23.14 & 12 & 2 & 10.18 & AEN\\
MASTER\,OT\,J072448.87+533952.1	& 07:24:48.86 & +53:39:51.49 & 10.6 & 0.1 & 0.42 & FRD \\
Gaia16bvf & 19:25:17.71 & +08:39:20.40 & 10 & 2 & 6.69 & AEN \\
V1454\,Cyg							               & 19:53:38.50 & +35:21:45.62 & 10.21 & 1.79 & 6.8 & AEN \\
WX\,LMi & 10:26:27.52 & +38:45:02.01 & 10.1 & 0.1 & 0.22 & DB \\
Gaia14abg & 17:30:47.93 & +50:00:16.65 & 9.5 & 0.1 & 0.0 & SWD \\
Ret1 & 03:34:34.43 & --64:00:57.88 & 9.35 & 0.02 & 0.14 & DB \\
BPM\,18764 & 08:02:00.41 & --53:27:49.36 & 9.2 & 0.1 & 0.3 & DB \\
OGLE-BLG-DN-0040 & 17:34:23.99 & --23:32:44.43 & 9 & 1 & 2.92 & SGD \\
ASASSN-14ib							            & 04:22:12.22 & --03:25:13.47 & 8.98 & 3.25 & 2.89 & AEN \\
MASTER\,OT\,J143453.02+023616.1	& 14:34:53.08 & +02:36:16.15 & 8.4 & 0.6 & 0.0 & FRD \\  
NSV\,15401							           & 01:55:10.12 & +69:42:40.14 & 8.32 & 0.08 & 0.13 & SWD \\ 
SBS\,1316+577A							    & 13:18:00.68 & +57:28:04.00 & 8.23 & 0.05 & 0.36 & FRD \\
SSS\,J035055.8-204817 & 03:50:56.02 & --20:48:15.95 & 7.98 & 0.06 & 0.0 & YSO \\
Larin\,2							                   & 12:48:50.77 & -41:26:54.65 & 7.95 & 0.13 & 0.32 & YSO \\ 
Gaia17cuc & 10:26:20.55	& --44:18:49.68	& 7.8 & 0.1 &	0.39 & FRD \\
ASASSN-17eo							            & 20:02:14.34 & +31:36:34.66 & 8 & 1 & 3.17 & AEN \\
SSS\,J162131.9--230140					& 16:21:31.92 & --23:01:40.73 & 7.3 & 0.1 & 0.59 & FRD\\
SSS\,J155929.1--223618					& 15:59:29.20 & -22:36:17.82 & 7.1 & 0.1 & 0.34 & YSO \\ 
MASTER\,OT\,J100950.32+471815.8	& 10:09:50.23 & +47:18:16.76 & 7.0 & 0.3 & 0.68 & FRD \\
SSS\,J155147.2--211323					& 15:51:47.08 & --21:13:23.86 & 6.89 & 0.13 & 0.61 & FRD \\
OGLE-GD-ECL-02234							& 10:45:49.79 & --61:29:57.04 & 6.79 & 0.08 & 0.53 & FRD \\
ASASSN-15ep							           & 08:21:06.24 & -72:20:12.09 & 6.2 & 0.1 & 0.71 & FRD \\
OGLE-BLG-DN-0428							   & 18:00:11.42 & --29:41:38.40 & 6 & 1 & 2.78 & SGD \\
MASTER\,OT\,J012916.47+321859.0	& 01:29:16.48 & +32:18:58.84 & 6 & 1 & 2.46 & FRD, AEN \\
NSV\,18024							           & 08:44:35.16 & --37:58:02.84 & 6.1 & 0.9 & 7.71 & AEN \\
MACHO\,401.48296.2600					&17:58:32.38 & --27:52:44.12 & 6 & 1 & 2.47 & AEN \\
Gaia17brd							               & 20:15:08.25 & +20:40:31.19 & 6 & 4 & 3.4 & AEN \\
MASTER\,OT\,J140957.49+290922.7	& 14:09:57.47 & +29:09:22.79 & 5.8 & 0.6 & 0.0 & FRD \\
V1419\,Aql							               & 19:13:06.79 & +01:34:23.24 & 6 & 2 & 5.41 & AEN \\
MASTER\,OT\,J120525.84+621743.3	& 12:05:25.88 & +62:17:43.04 & 5.7 & 0.5 & 2.23 & FRD, AEN \\
MASTER\,OT\,J020404.19+741804.6	& 02:04:03.60 & +74:18:02.45 & 5.6 & 1.3 & 4.56 & AEN \\ 
Gaia17cva							               & 19:45:37.72 & +28:05:32.88 & 6 & 2 & 1.46 & SGD \\
OGLE-BLG-DN-0266							    & 17:54:53.98 & --21:22:40.19 & 5 & 2 & 3.86 & AEN \\
Gaia17aoi						 	                   & 13:24:44.33 & --14:23:35.65 & 5 & 1 & 2.45 & AEN \\
OGLE-BLG-DN-0824							    & 18:10:04.90 & --29:05:23.58 & 5 & 1 & 3.61 & SGD, AEN \\
CSS\,111021:220328+141059			& 22:03:28.11 & +14:11:00.49 & 4 & 1 & 0.0 & SGD \\
OGLE-BLG-DN-0011							    & 17:17:26.01 & --28:33:23.79 & 4 & 1 & 4.47 & SGD, AEN \\
MASTER\,OT\,J203824.15+174242.3	& 20:38:24.10 & +17:42:43.15 & 4 & 2 & 2.98 & AEN \\
OGLE-BLG-DN-0087							    & 17:43:07.76 & --34:19:28.10 & 3 & 1 & 3.2 & SGD, AEN \\
OGLE-BLG-DN-0156							    & 17:49:49.23 & --21:22:13.57 & 3 & 2 & 5.32 & SGD, AEN \\
CSS\,150422:172535+231215			& 17:25:34.90 & +23:12:14.33 & 3 & 3 & 5.14 & AEN \\
MASTER\,OT\,J051042.59+513540.0	& 05:10:42.60 & +51:35:39.82 & 3 & 1 & 3.14 & AEN \\
CSS\,081201:213947+170658			& 21:39:47.16 & +17:06:56.53 & 3 & 2 & 1.65 & SGD \\
Gaia16bfi & 16:37:08.62	& --67:36:56.46 & 5.1 & 1.9 & 4.70 & AEN \\
ASASSN-13cv	& 22:10:25.35 & +30:46:10.06 & 4.5 & 0.8 & 	1.19 & FRD \\
OGLE-BLG-DN-0584 & 18:03:46.14 & --27:15:33.70 & 4.2 & 0.8 & 1.74 & FRD \\
ASASSN-17gc	& 19:51:36.94 & --00:59:04.05 &	4 &	2 &	0.71 & SGD \\
ASASSN-16cd	& 19:06:38.16 & +33:09:03.17 &	3.4 & 1.5 & 5.5 & SGD, AEN \\
ASASSN-17nm	& 09:46:09.11 & --57:14:20.41 &	3.1 & 1.2 &	0.0 & SGD \\
Gaia17aok & 21:52:55.72	& 59:18:18.71 & 3.1 & 1.2 & 0.0 & SGD \\
CSS\,120313:131043--042600			& 13:10:42.68 & --04:26:00.73 & 3 & 2 & 2.2 & AEN \\
OGLE-BLG-DN-0216 & 17:52:24.02 & --32:16:51.54 & 2.2 & 1.5 & 2.7 & SGD, AEN \\
EL\,Aql							                      & 18:56:01.87 & -03:19:18.80 & 2 & 3 & 2.68 & AEN \\
MASTER\,OT\,J182201.93+324906.7	& 18:22:01.80 & +32:49:00.57 & 2.0 & 	2.0 & 4.85 & SGD, AEN \\
MLS\,101009:010045+010019			& 01:00:44.71 & +01:00:18.48 & 2 & 2 & 1.53 & Q \\
EU\,Cnc							                   & 08:51:27.17 & +11:46:56.94 & 2 & 2 & 0.0 & SGD\\
OGLE-BLG-DN-0181							    & 17:51:01.18 & --29:14:38.30 & 2 & 2 & 2.82 & AEN \\
Gaia17afs & 17:35:17.86 & +01:32:48.79 & 1.1 & 1.9 & 3.0 & SGD, AEN \\
OGLE-BLG-DN-0054							    & 17:38:24.00 & --21:54:26.87 & 2 & 2 & 2.13 & AEN \\
DO\,Vul							                   & 19:52:10.72 & +19:34:42.14 & 0.8 & 2 & 4.78 & AEN \\
Gaia17bqf							               & 19:59:19.93 & +16:24:40.31 & --0.2 & 2 & 3.07 & AEN \\
V1722\,Aql & 19:14:09.74 & +15:16:38.25 & --1 & 3 &	3.07 & SGD, AEN \\ 
\end{longtable}

\setlength{\tabcolsep}{0.1cm}
\begin{table*}
\caption{CV and CV candidates with parallaxes $\varpi + 3\sigma_\varpi >= 6.66\,\mathrm{mas}$ that are located further than 150~pc.}\label{tab:cv_excluded}
\begin{tabular}{lcccccc}
 \toprule
 System         & $\alpha$ & $\delta$ & $\varpi$ & $\sigma_\varpi$ & Distance & $P(d < 150$\,pc)\\
                     &               &               &  (mas)    &   (mas)               &  (pc)       &     (\%)  \\
\midrule
V1108\,Her							              & 18:39:26.14 & +26:04:09.96 & 6.6 & 0.1 & $152 \pm 3$ & 0.24 \\
\addlinespace[0.05cm]
EK\,TrA							                  & 15:14:00.10 & --65:05:36.65 & 6.58 & 0.04 & $152 \pm 1$ & 0.02 \\
\addlinespace[0.05cm]
BZ\,UMa							                  & 08:53:44.22 & +57:48:40.35 & 6.56 & 0.06 & $153 \pm 1$ &	 0.04 \\
\addlinespace[0.05cm]
EI\,Psc							                      & 23:29:54.17 & +06:28:12.11 & 6.55 & 0.07 & $153 \pm 2$ & 0.04 \\
\addlinespace[0.05cm]
MASTER\,OT\,J050806.84+712352.0	& 05:08:06.78 & +71:23:51.69 & 6.5 & 2.8 & $487^{+489}_{-255}$ & 0.04 \\
\addlinespace[0.1cm]
FL\,Psc							                      & 00:25:11.04 & +12:17:11.81 & 6.5 & 0.1 & $154 \pm 3$ & 0.11\\
\addlinespace[0.05cm]
V348\,Pav							              & 19:56:48.05 & --60:34:30.00 & 6.48 & 0.08 & $154 \pm 2$	& 0.006 \\
\addlinespace[0.05cm]
ASASSN-16jg							           & 14:45:35.72 & --39:20:26.73 & 6.4 & 0.3 & $158^{+9}_{-8}$ & 0.17 \\
\addlinespace[0.05cm]
1RXS\,J083842.1--282723			    & 08:38:43.33 & --28:27:00.95 & 6.4 & 0.1 & $157 \pm 3$ & 0.004 \\
\addlinespace[0.05cm]
SDSS\,J150551.58+065948.7			& 15:05:51.61 & +06:59:48.49 & 6.3 & 0.4 & $162^{+13}_{-11}$ & 0.14 \\
\addlinespace[0.05cm]
EF\,Eri							                      & 03:14:13.41 & --22:35:43.77 & 6.3 & 0.3 & $161 \pm 7$ & 0.05 \\
\addlinespace[0.05cm]
MASTER\,OT\,J220559.40--341434.9	& 22:05:59.47 & --34:14:34.15 & 6.2 & 0.3 & $161 \pm 7$ &  0.04 \\
\addlinespace[0.05cm]
CSS\,081221:050716+125314			& 05:07:16.24 & +12:53:14.16 & 5.4 & 2.3 & $485^{+467}_{-237}$ & 0.02 \\
\addlinespace[0.05cm]
ASASSN-18bh							           & 01:09:52.87 & +47:57:11.11 & 5.2 & 1.7 & $361^{+354}_{-146}$ & 0.02 \\
\addlinespace[0.05cm]
ASASSN-14ip							           & 20:50:23.43 & --48:37:13.71 & 5.0 & 1.1 & $253^{+115}_{-60}$ &  0.01\\
\addlinespace[0.05cm]
V498\,Hya							               & 08:45:55.06 & +03:39:29.28 & 4.9 & 2.0 & $469^{+446}_{-217}$ & 0.01\\
\addlinespace[0.05cm]
ASASSN-15td							           & 12:15:13.70 & --01:46:41.56 & 4.9 & 2.3 & $527^{+478}_{-255}$ & 0.01\\
\addlinespace[0.05cm]
OGLE-BLG-DN-0183							    & 17:51:05.70 & --28:03:37.84 & 4.8 & 1.8 & $440^{+416}_{-193}$ & 0.01\\
\addlinespace[0.05cm]
ASASSN-16ee							           & 08:35:42.43 & --31:21:47.92 & 4.5 & 1.3 & $335^{+254}_{-110}$ & 0.006\\
\addlinespace[0.05cm]
CSS 170417:080539+354055			& 08:05:38.98 & +35:40:54.94 & 4.3 & 1.7 & $461^{+413}_{-196}$ & 0.006\\ 
\addlinespace[0.05cm]
ASASSN-13bd							           & 23:59:58.00 & --12:54:32.68 & 4.3 & 2.2 & $575^{+485}_{-271}$ & 0.007\\ 
\addlinespace[0.05cm]
SDSS\,J125641.29--015852.0          & 12:56:41.29 & --01:58:51.74 & 4.0 & 1.1 & $350^{+202}_{-96}$ & 0.0007\\ 
\addlinespace[0.05cm]
CSS\,080927:212522--102627        & 21:25:21.76 & --10:26:28.18 & 4.0 & 1.6 & $471^{+404}_{-194}$ & 0.003\\ 
\addlinespace[0.05cm]
MASTER\,OT\,J122126.39--311248.3 & 12:21:26.40 & --31:12:48.49 & 3.8 & 1.0 & $363^{+217}_{-104}$ & 0.0002\\ 
\addlinespace[0.05cm]
CSS\,111103:074400+415504        & 07:44:00.47 & +41:55:03.56 & 3.7 & 2.4 & $636^{+499}_{-295}$ & 0.004\\ 
\addlinespace[0.05cm]
IK\,Leo	                                            & 10:21:46.45 & +23:49:25.91 & 3.6 & 1.4 & $499^{+402}_{-198}$ & 0.0008\\  
\addlinespace[0.05cm]
ASASSN-15rj                                    & 02:59:38.35 & +44:57:04.77 & 3.6 & 1.6 & $529^{+429}_{-220}$ & 0.001\\   
\addlinespace[0.05cm]
MASTER\,OT\,J070740.72+702630.0 & 07:07:40.55 & +70:26:30.30 & 3.5 & 1.1 & $416^{+287}_{-135}$ & 0.0002\\   
\addlinespace[0.05cm]
ASASSN-15ef                                   & 16:49:40.59 & --17:50:09.72 & 3.5 & 1.6 & $549^{+444}_{-234}$ & 0.001\\   
\addlinespace[0.05cm]
CSS\,110124:032934+182530         & 03:29:33.92 & +18:25:29.57 & 3.5 & 1.1 & $406^{+265}_{-126}$ & 0.0001\\ 
\addlinespace[0.05cm]
ASASSN-17jf                                    & 20:29:17.10 & --43:40:19.18 & 3.5 & 1.1 & $426^{+297}_{-140}$ & 0.0001\\ 
\addlinespace[0.05cm]
ASASSN-17mw	                                 & 02:49:07.33 & +48:51:01.16 & 3.5 & 1.1 & $408^{+265}_{-127}$ & 0.00009 \\ 
\addlinespace[0.05cm]
Gaia16apf	                                        & 00:34:33.39 & +54:28:42.04 & 3.5 & 1.3 & $488^{+381}_{-186}$ & 0.0004\\ 
\addlinespace[0.05cm]
ASASSN-15gm	                                 & 19:37:13.59 & --22:57:06.31 & 3.5 & 1.6 & $544^{+435}_{-226}$ & 0.0009\\ 
\addlinespace[0.05cm]
ASASSN-16jb							          & 17:50:44.96 & --25:58:37.45 & 3.42 & 1.2 & $456^{+334}_{-160}$ & 0.0002\\ 
\addlinespace[0.05cm]
ASASSN-16do							          & 06:34:12.71 & --32:59:49.49 & 3.1 & 1.3 & $525^{+392}_{-198}$ & 0.0001\\ 
\addlinespace[0.05cm]
ASASSN-13ck							          & 00:11:33.73 & +04:51:22.43 & 3.1 & 1.6 & $590^{+452}_{-247}$ & 0.0005\\  
\addlinespace[0.05cm]
ASASSN-15px							          & 23:08:57.87 & --65:59:32.49 & 3.1 & 1.3 & $547^{+413}_{-214}$ & 0.0002\\   
\addlinespace[0.05cm]
CSS\,110406:152159+261223       &15:21:58.84 & +26:12:23.30 & 2.7 & 2.1 & $689^{+499}_{-303}$ & 0.001 \\    
\addlinespace[0.05cm]
CSS\,101108:022436+372021       & 02:24:36.44 & +37:20:21.40 & 2.7 & 1.8 & $671^{+485}_{-287}$ & 0.0005 \\     
\addlinespace[0.05cm]
KK\,Cnc							                  & 08:07:14.25 & +11:38:12.32 & 2.5 & 1.6 & $670^{+476}_{-278}$ & 0.0002 \\
\addlinespace[0.05cm]
ASASSN-16jk							          &15:40:24.84 & +23:07:50.86 & 2.5 & 1.4 & $655^{+457}_{-260}$ & 0.00004 \\
\addlinespace[0.05cm]
MASTER\,OT\,J152701.21--314433.6 &	15:27:01.25 & --31:44:35.30 & 2.5 & 1.5 & $667^{+468}_{-270}$ & 0.00008 \\
\addlinespace[0.05cm]
ASASSN-15aw							           & 01:57:46.15 & +51:10:23.88 & 2.4 & 1.7 & $695^{+487}_{-292}$ & 0.0002  \\
\addlinespace[0.05cm]
CSS\,100108:081031+002429        & 08:10:30.61 & +00:24:28.32 & 2.4 & 2.2 & $716^{+504}_{-312}$ & 0.0008   \\
\addlinespace[0.05cm]
ASASSN-15nf	                                 & 20:12:42.71 & +15:44:44.92 & 2.4 & 2.4 & $724^{+509}_{-319}$ & 0.001 \\
\addlinespace[0.05cm]
CSS\,110430:091710+314309	       & 09:17:09.87 & +31:43:07.59 & 2.3 & 1.5 & $686^{+472}_{-376}$ & 0.00005 \\
\addlinespace[0.05cm]
ASASSN-17bi                                   & 02:16:05.42 & +68:39:03.61 & 2.3 & 1.5 & $694^{+477}_{-282}$ & 0.00007 \\
\addlinespace[0.05cm]
CSS\,090928:030141+241541        & 03:01:40.52 & +24:15:41.35 & 2.3 & 2.5 & $732^{+511}_{-323}$ & 0.002 \\
\addlinespace[0.05cm]
ASASSN-14kk                                   & 01:32:02.78 & --10:43:57.72 & 2.3 & 1.5 & $689^{+471}_{-276}$ & 0.00004 \\
\addlinespace[0.05cm]
MASTER\,OT\,J211855.08+280314.9 & 21:18:55.10 & +28:03:15.39 & 2.2 & 3.1 & $744^{+520}_{-335}$ & 0.002 \\
\addlinespace[0.05cm]
SSS\,110125:103550--424610         & 10:35:49.64 & --42:46:10.14 & 2.1 & 2.2 & $741^{+508}_{-320}$ & 0.0006 \\
\addlinespace[0.05cm]
CSS\,150822:232026+221833			& 23:20:26.23 & +22:18:34.05 & 0.3 & 3.0 & $814^{+523}_{-346}$ & 0.0007 \\
\addlinespace[0.05cm]
CSS\,090926:230711+294010			& 23:07:11.34 & +29:40:11.33 & 0.2 & 2.2 & $857^{+519}_{-344}$ & 0.00007 \\ 
\bottomrule
\end{tabular}
\end{table*}

\setlength{\tabcolsep}{0.15cm}
\setlength\cellspacetoplimit{3pt}
\setlength\cellspacebottomlimit{3pt}
\begin{table*}
\begin{adjustwidth}{-2.2em}{0em}
\caption{CVs with pre-\gaia distance estimates $d\le150$\,pc.}\label{tab:dist_pre-gaia}
\begin{tabular}{@{}llccc Sc | Sc ccccc@{}}
 \toprule
               &          &          &                  &          &        &           &   \multicolumn{4}{c}{\gaia}\\
System         & \multicolumn{1}{c}{\gaia DR2 ID} & $P_\mathrm{orb}$ & Distance & Method & Reference &  $G$ & $G_\mathrm{BP}$ & $G_\mathrm{RP}$ & $\varpi$ & $\sigma_\varpi$ & Distance\\
                         &          &  (min)           &  (pc)    &        &  &  (mag) & (mag) & (mag) &  (mas)   &   (mas)         &  (pc)   \\
\midrule 
WZ\,Sge & 1809844934461976832 & 81.63 & 43.30$^{+1.60}_{-1.50}$ & a & 1 & 15.21 & 15.21 & 15.06 & 22.16 &  0.04 &  45.13$\pm 0.08$ \\  
AY\,Lyr & 2096934223687181696 & 105.55 & 52 & c & 2 &  17.93 & 17.96 & 17.50 & 2.22 &  0.13 &  452$^{+32}_{-22}$ \\  
XZ\,Eri & 5097770801875122432 & 88.7 & 66 & c & 2 & 19.25 & 19.28 & 19.00 & 3.0 & 0.3 & 331$^{+44}_{-24}$\\ 
VW\,Hyi & 4653893040002306432 & 106.95 & 64$^{+20} _{-17}$ & e & 3 & 13.84 & 13.94 & 14.45 & 18.53 &  0.02 &  53.96$\pm 0.06 $ \\  
EX\,Hya & 6185040879503491584 & 98.26 & 64.5$\pm 1.2$ & a & 4 & 13.21 & 13.23 & 12.88 & 17.56 &  0.04 &  56.95$\pm 0.13$ \\  
GD\,552 & 2208124536065383424 & 102.73 & 74$\pm 4$ & e & 5 & 16.46 & 16.46 & 16.18 & 12.35 &  0.05 &  81.0$\pm 0.3$ \\  
QZ\,Vir & 3800596876396315648 & 84.70 & 76 & c & 2 & 16.06 & 16.12 & 15.76 & 7.81 &  0.07 & $128 \pm 1$ \\
AM\,Her & 2123837555230207744 & 185.65 & 79$^{+ 8} _{- 6}$ & a & 1 & 13.58 & 13.86 & 12.85 & 11.40 &  0.02 &  87.76$\pm 0.14$ \\  
1RXS J105010.3-140431 & 3750072904055666176 & 88.56 & 80$ \pm 20$ & e & 5  &  17.17 & 17.21 & 17.08 & 9.14 &  0.11 &  109$\pm 1$ \\  
AR\,UMa & 783921244796958208 & 115.92 & 86$^{+10} _{- 8}$ & a & 6 &  16.26 & 16.35 & 15.78 & 9.87 &  0.12 &  101$\pm 1$ \\  
V455\,And & 1920126431748251776 & 81.08 & 90$\pm 15$ & b & 7 & 16.06 & 16.13 & 15.71 & 13.24 &  0.06 &  75.5$\pm 0.3$ \\  
ASASSN-14ag & 3071240270519385856 & 86.85 & 90 & e &  8  & 16.18 & 16.17 & 15.70 & 5.63 &  0.09 &  178$ \pm 3$ \\  
IX\,Vel* & 5515820034889610112 & 279.25 & 96$\pm 1$ & a & 9 & 9.32 &  9.34 &  9.27 & 11.04 &  0.03 &  90.6$\pm 0.2$ \\   
VY\,Aqr & 6896767366186700416 & 90.85 & 97$^{+15}_{-12}$ & a & 1 &  16.86 & 16.96 & 16.44 & 7.24 &  0.14 &  138$\pm 3$ \\   
U\,Gem & 674214551557961984 & 254.74 & 97$\pm 7$ & c & 10  & 13.91 & 14.38 & 13.11 & 10.71 &  0.03 &  93.3$\pm 0.3$ \\   
IGR\,J18308--1232 & 4153024090088033280 & -- & 100 & c & 11 &  17.65 & 18.07 & 16.89 & 0.49 &  0.15 &  1595$^{+514}_{-193}$  \\  
V426\,Oph & 4471872295941149056 & 410.83 & 100 & e & 12 & 12.37 & 12.65 & 11.77 & 5.20 &  0.04 &  192.5$\pm 1.5$ \\   
WW\,Cet & 2427474150870397056 & 253.15 & 100 & c & 2 & 13.55 & 13.65 & 12.96 & 4.59 & 0.05 & 218$\pm 2$\\   
OY\,Car & 5242787486412627072 & 90.89 & 100 & c & 13 & 15.62 & 15.64 & 15.21 & 11.01 &  0.03 &  90.8$\pm 0.2$ \\  
V2051\,Oph & 4111991385628196224 & 89.90 & 102 $\pm 16$ & e & 5 &  15.37 & 15.46 & 14.87 & 8.90 &  0.07 &  112.4$\pm 0.9$ \\  
AE\,Aqr* & 4226332451596335616 & 592.78 & 102$^{+42}_{-23}$ & a & 1  & 10.95 & 11.47 & 10.26 & 10.97 &  0.06 &  91.2$\pm 0.5$ \\  
GW\,Lib & 6226943645600487552 & 76.78 & 104$^{+30}_{-20}$ & a & 1 & 16.49 & 16.49 & 16.32 &  8.87 &  0.08 &  113$\pm 1$ \\  
DI\,UMa & 1013298268207936128 & 78.59 & 107 & c & 2 & 17.75 & 16.86 & 16.82 & 1.46 &  0.08 &  685$^{+43}_{-31}$ \\   
V3885\,Sgr* & 6688624794231054976 & 298.31 & 110$\pm 30$ & a & 14 & 10.25 & 10.28 & 10.16 & 7.54 &  0.08 &  133$\pm 1$ \\   
HU\,Aqr & 6911950900211768704 & 125.02 & 111 & c & 2 & 16.47 & 16.66 & 15.88 & 5.20 &  0.06 &  192$\pm 2$ \\  
Z\,Cha & 5210507882302442368 & 107.28 & 112$\pm 8$ & c & 10 & 15.85 & 15.94 & 15.19 & 8.66 &  0.12 &  115$\pm 1$ \\   
AH\,Eri & 3176908972944418816 & 344.30 & 113 & c & 2 &  17.46 & 17.85 & 16.75 & 0.82 &  0.09 &  1129$^{+134}_{-80}$ \\    
EF\,Eri & 5099482805904892288 & 81.02 & 113$^{+19}_{-16}$ & a  & 1 & 18.21 & 18.17 & 18.11 & 6.3 & 0.3 & $161 \pm 7$\\    
BK\,Lyn & 702296666944246784 & 107.97 & 114 & c & 2 & 14.52 & 14.48 & 14.45 & 1.98 &  0.07 &  505$^{+20}_{-16}$ \\   
SS\,Cyg* & 1972957892448494592 & 396.19 & 114$\pm 2$ & a & 15 & 11.69 & 12.11 & 10.95 &  8.72 &  0.05 &  114.6$\pm 0.6$ \\   
ST\,LMi & 3996419759863758592 & 113.89 & 115$^{+21}_{-22}$ & b & 16 &  16.13 &   -- &   -- & 8.83 &  0.08 &  113$\pm 1$ \\ 
DH\,Aql & 4200218019655998720 & -- & 116 & c & 2 &  17.97 & 18.28 & 17.43 & 3.6 &  0.2 &  282$^{+18}_{-13}$ \\  
IQ\,Eri & 5078976609103251456 & -- & 116$^{+116}_{-58}$ & e & 3 & 17.18 & 17.11 & 16.85 & 5.19 &  0.17 &  193$^{+ 7}_{-6}$ \\  
IP\,Peg & 2824150286583562496 & 227.82 & 124 & e & 12 & 14.71 & 15.27 & 13.88 & 7.08 &  0.05 &  141$\pm 1$ \\   
VV\,Pup & 5719598950133755392 & 100.44 & 124$^{+17}_{-14}$ & a & 6 & 15.93 & 16.04 & 15.48 & 7.30 &  0.05 & 137$\pm 0.9$ \\   
MR\,Ser & 1203639265875666304 & 113.47 & 126$^{+14}_{-12}$ & a & 6 & 16.23 & 16.47 & 15.64 & 7.59 &  0.05 & 131.8$\pm 0.9$ \\  
TV\,Col & 2901783160488793728 & 329.18 & 128$\pm 1$ & a & 17 & 13.93 & 14.03 & 13.61 & 1.95 &  0.02 &  512$\pm 4$ \\   
IGR\,J17195--4100 & 5959894875620104064 & 240.34 & 130 & c & 11 &  15.33 & 15.55 & 14.75 & 1.53 &  0.04 &  653$^{+30} _{-27}$ \\  
EQ\,Cet & 5041907811522399488 & 92.82 & 130 & c & 18 & 17.34 & 17.41 & 16.78 &  3.51 &  0.11 &  285$^{+ 10} _{- 8}$ \\  
BL\,Hyi & 4697621824327141248 & 113.64 & 130 & c & 19 & 17.25 & 17.45 & 16.70 & 7.65 &  0.07 & $131 \pm 1$ \\
BW\,Scl & 2307289214897332480 & 78.23 & 131$\pm 18$ & b & 20 & 16.26 & 16.26 & 16.10 & 10.60 &  0.10 &  94.4$\pm 0.9$ \\  
HT\,Cas & 426306363477869696 & 106.05 & $131^{+22}_{-17}$ & a & 8 & 16.35 & 16.48 & 15.80 & 7.07 & 0.06 & $141 \pm     1$ \\
RX\,And & 374510294830244992 & 302.24 & 135 & e & 12 & 13.17 & 13.38 & 12.50 & 5.03 & 0.05 & 199$\pm 2$ \\
V406\,Vir & 3681313024562519552 & 80.52 & 140$\pm 35$ & e & 5 & 17.72 & 17.71 & 17.55 &  5.91 & 0.16 &  169$\pm 5$ \\ 
V834\,Cen & 6096905573613586944 & 101.52 & 144$^{+18} _{-23}$ & b & 16 & 16.66 & 16.82 & 16.07 & 8.9 &  0.2 &  113$\pm 3$ \\  
V405\,Peg & 2838503311371673472 & 255.81 & 149 & a & 21 & 15.28 & 16.05 & 14.29 & 5.78 &  0.06 &  173$\pm 2$ \\  
V2301\,Oph & 4476137370261520000 & 112.97 & 150 & b & 22 & 16.75 & 16.94 & 15.86 & 8.24 &  0.08 &  121$\pm 1$ \\  
SX\,LMi & 733329416268149376 & 96.72 & 150 & c & 2 & 16.69 & 16.77 & 16.30 & 3.08 &  0.12 &  325$^{+14} _{-11}$ \\  
CU\,Vel & 5524430207364715520 & 113.04 & 150$\pm 50$ & b & 23 & 16.71 & 16.93 & 16.15 & 6.29 &  0.06 &  159$\pm 2$ \\  
\bottomrule 
\end{tabular}
\begin{tablenotes}
\item \textbf{Notes.}
Systems highlighted with a star have a parallax measurement from \gaia DR1 \citep{Ramsay+2017} but we do not report them in this table since they do not represent a pre-\gaia distance determination. The method abbreviations recall the list in Section~\ref{subsec:completeness}. The distances reported in the last column have been determined from the \gaia parallaxes as described in Section~\ref{sec:150pc}.
\item \textbf{References:} (1) \citet{Thorstensen2003} , (2) \citet{Sproats+1996}, (3) \citet{Pretorius+2012}, (4) \citet{Beuermann+2003}, (5) \citet{Patterson2011}, (6) \citet{Thorstensen+2008}, (7) \citet{Araujo-Betancor+2005}, (8) \citet{Thorstensen+2016}, (9) \citet{Linnell+2007}, (10) \citet{Beuermann2006}, (11) \citet{Bernardini+2012}, (12) \citet{Warner1987}, (13) \citet{Sherrington+1982}, (14) \citet{Linnell+2009}, (15) \citet{Miller-Jones+2013}, (16) \citet{Araujo-Betancor+2005a}, (17) \citet{McArthur+2001}, (18) \citet{Schwope+1999}, (19) \citet{Beuermann+1985}, (20) \citet{Gaensicke+2005}, (21) \citet{Thorstensen+2009}, (22) \citet{Szkody+1996}, (23) \citet{Boris+1999}.
\end{tablenotes}
\end{adjustwidth}
\end{table*}


\bsp	
\label{lastpage}
\end{document}